\input harvmac
\input epsf

\newcount\figno
\figno=0
\def\fig#1#2#3{
\par\begingroup\parindent=0pt\leftskip=1cm\rightskip=1cm\parindent=0pt
\baselineskip=12pt
\global\advance\figno by 1
\midinsert
\epsfxsize=#3
\centerline{\epsfbox{#2}}
\vskip 14pt

{\bf Fig. \the\figno:} #1\par
\endinsert\endgroup\par
}
\def\figlabel#1{\xdef#1{\the\figno}}
\def\encadremath#1{\vbox{\hrule\hbox{\vrule\kern8pt\vbox{\kern8pt
\hbox{$\displaystyle #1$}\kern8pt}
\kern8pt\vrule}\hrule}}

\overfullrule=0pt

\noblackbox
\parskip=1.5mm
%\def\semi{;~}

%%%%%%%%%%%%%%%%%%%%%%%%%%%%%%%%%%%%%%%%%%%%%%%%%%%%%%%%%%%%%%%%%%%
%%%  modify title page
%%%%%%%%%%%%%%%%%%%%%%%%%%%%%%%%%%%%%%%%%%%%%%%%%%%%%%%%%%%%%%%%%%%
\def\Title#1#2{\rightline{#1}\ifx\answ\bigans\nopagenumbers\pageno0
%   \vskip0.5in
\else\pageno1\vskip.5in\fi \centerline{\titlefont #2}\vskip .3in}

\font\caps=cmcsc10
%\def\listrefs{\footatend\bigskip\bigskip\immediate\closeout\rfile
%\writestoppt \baselineskip =11pt\centerline{{\secfont References}}
%\bigskip{\frenchspacing\parindent =20pt \escapechar +'
%\input\jobname.refs \vfill\eject}\nonfrenchspacing}
%%%%%%%%%%%%%%%%%%%%%%%%%%%%%%%%%%%%%%%%%%%%%%%%%%%%%%%%%%%%%%%%%%%%%%%%%%%%

\noblackbox
\parskip=1.5mm
%\def\semi{;~}

%%%%%%%%%%%%%%%%%%%%%%%%%%%%%%%%%%%%%%%%%%%%%%%%%%%%%%%%%%%%%%%%%%%%%

%%%%%%%%%%%%%%%%%%%%%%%%%%%%%%%%%%%%%%%%%%%%%%%%%%%%%%%%%%%%%%%%%%%%%
%%%%%%%%%%%%%%%%%%%%    some definitions    %%%%%%%%%%%%%%%%%%%%%%%%%
%%%%%%%%%%%%%%%%%%%%%%%%%%%%%%%%%%%%%%%%%%%%%%%%%%%%%%%%%%%%%%%%%%%%%

            \def\CZ{{\cal Z}}
   
 \def\CH{{\cal H}} \def\CI{{\cal I}} 
 \def\CR{{\cal R}}

%%%%%%%%%%%%%%%%%%%%%%%%%%%%%%%%%%%%%%%%%%%%%%%%%%%%%%%%%%%%%%%%%%%%%

\def\dj{\hbox{d\kern-0.347em \vrule width 0.3em height 1.252ex depth
-1.21ex \kern 0.051em}}

\def\half{{1\over 2}\,}

\def\Tr{{\rm Tr\,}}
\def\tr{{\rm tr\,}}

\def\ket{\rangle}

\def\pt{\partial}

\def\Dirac{\,\raise.15ex\hbox{/}\mkern-13.5mu D}
\def\dirac{\,\raise.15ex\hbox{/}\kern-.57em \partial}
\def\aslash{\,\raise.15ex\hbox{/}\mkern-13.5mu A}

\def\shalf{{\ifinner {\textstyle {1 \over 2}}\else {1 \over 2} \fi}}
\def\sshalf{{\ifinner {\scriptstyle {1 \over 2}}\else {1 \over 2} \fi}}
\def\sfourth{{\ifinner {\textstyle {1 \over 4}}\else {1 \over 4} \fi}}
%%%%%%%%%%%%%%%%%%%%%%%%%%%%%%%%%%%%%%%%%%%%%%%%%%%%%%%%%%%%%%%%%%%%%%
%%%%%%%%%%%%%%%%%%%%%%%        references         %%%%%%%%%%%%%%%%%%%%%%
%%%%%%%%%%%%%%%%%%%%%%%%%%%%%%%%%%%%%%%%%%%%%%%%%%%%%%%%%%%%%%%%%%%%%%%
%%%%%%%%%%%%%%%%%%%
%%%

\lref\rtouring{
  J.~L.~F.~Barbon and E.~Rabinovici,
  ``Touring the Hagedorn ridge,''
  arXiv:hep-th/0407236.
  %%CITATION = HEP-TH/0407236;%%
  }
  
  \lref\rbek{ J.~L.~F.~Barbon and E.~Rabinovici,
  ``Remarks on black hole instabilities and closed string tachyons,''
  Found.\ Phys.\  {\bf 33}, 145 (2003)
  [arXiv:hep-th/0211212].
  %%CITATION = FNDPA,33,145;%%
  }
  
  \lref\rlstbasic{
  N.~Seiberg,
  ``New theories in six dimensions and matrix description of M-theory on  T**5
  and T**5/Z(2),''
  Phys.\ Lett.\  B {\bf 408}, 98 (1997)
  [arXiv:hep-th/9705221].
  O.~Aharony, M.~Berkooz, S.~Kachru, N.~Seiberg and E.~Silverstein,
  ``Matrix description of interacting theories in six dimensions,''
  Adv.\ Theor.\ Math.\ Phys.\  {\bf 1}, 148 (1998)
  [arXiv:hep-th/9707079].
  M.~Berkooz, M.~Rozali and N.~Seiberg,
  ``Matrix description of M theory on T**4 and T**5,''
  Phys.\ Lett.\  B {\bf 408}, 105 (1997)
  [arXiv:hep-th/9704089].}
  
  \lref\raharonyrew{
  O.~Aharony,
  ``A brief review of 'little string theories',''
  Class.\ Quant.\ Grav.\  {\bf 17}, 929 (2000)
  [arXiv:hep-th/9911147].}
  
  \lref\rkut{
  D.~Kutasov and D.~A.~Sahakyan,
  ``Comments on the thermodynamics of little string theory,''
  JHEP {\bf 0102}, 021 (2001)
  [arXiv:hep-th/0012258].
  %%CITATION = JHEPA,0102,021;%%
  }
  
  \lref\rmaldafiveb{
  J.~M.~Maldacena,
  ``Statistical Entropy of Near Extremal Five-branes,''
  Nucl.\ Phys.\  B {\bf 477}, 168 (1996)
  [arXiv:hep-th/9605016].
  %%CITATION = NUPHA,B477,168;%%
  }
  
  \lref\radscft{
  J.~M.~Maldacena,
  ``The large N limit of superconformal field theories and supergravity,''
  Adv.\ Theor.\ Math.\ Phys.\  {\bf 2}, 231 (1998)
  [Int.\ J.\ Theor.\ Phys.\  {\bf 38}, 1113 (1999)]
  [arXiv:hep-th/9711200].
 S.~S.~Gubser, I.~R.~Klebanov and A.~M.~Polyakov,
  ``Gauge theory correlators from non-critical string theory,''
  Phys.\ Lett.\  B {\bf 428}, 105 (1998)
  [arXiv:hep-th/9802109].
 E.~Witten,
  ``Anti-de Sitter space and holography,''
  Adv.\ Theor.\ Math.\ Phys.\  {\bf 2}, 253 (1998)
  [arXiv:hep-th/9802150].
  }
  
  \lref\ritzhaki{
  N.~Itzhaki, J.~M.~Maldacena, J.~Sonnenschein and S.~Yankielowicz,
  ``Supergravity and the large N limit of theories with sixteen
  supercharges,''
  Phys.\ Rev.\  D {\bf 58}, 046004 (1998)
  [arXiv:hep-th/9802042].
  %%CITATION = PHRVA,D58,046004;%%
  }
  
  \lref\rminwallasei{
  S.~Minwalla and N.~Seiberg,
  ``Comments on the IIA NS5-brane,''
  JHEP {\bf 9906}, 007 (1999)
  [arXiv:hep-th/9904142].
  %%CITATION = JHEPA,9906,007;%%
  }

  \lref\rlszlst{
   O.~Aharony, A.~Giveon and D.~Kutasov,
  ``LSZ in LST,''
  Nucl.\ Phys.\  B {\bf 691}, 3 (2004)
  [arXiv:hep-th/0404016].
  %%CITATION = NUPHA,B691,3;%%
}

\lref\rfivebraneholo{
O.~Aharony, M.~Berkooz, D.~Kutasov and N.~Seiberg,
  ``Linear dilatons, NS5-branes and holography,''
  JHEP {\bf 9810}, 004 (1998)
  [arXiv:hep-th/9808149].
  %%CITATION = JHEPA,9810,004;%%
}

\lref\rpeetpol{
A.~W.~Peet and J.~Polchinski,
  ``UV/IR relations in AdS dynamics,''
  Phys.\ Rev.\  D {\bf 59}, 065011 (1999)
  [arXiv:hep-th/9809022].}
  
\lref\rcoiled{
 M.~Berkooz and M.~Rozali,
  ``Near Hagedorn dynamics of NS fivebranes, or a new universality class  of
  coiled strings,''
  JHEP {\bf 0005}, 040 (2000)
  [arXiv:hep-th/0005047].
  %%CITATION = JHEPA,0005,040;%%
  }
  
  \lref\rdlcq{
  O.~Aharony, M.~Berkooz and N.~Seiberg,
  ``Light-cone description of (2,0) superconformal theories in six
  dimensions,''
  Adv.\ Theor.\ Math.\ Phys.\  {\bf 2}, 119 (1998)
  [arXiv:hep-th/9712117].
  %%CITATION = 00203,2,119;%%
  }
  
  \lref\rdlcqt{
  O.~Aharony and M.~Berkooz,
  ``IR dynamics of d = 2, N = (4,4) gauge theories and DLCQ of 'little  string
  theories',''
  JHEP {\bf 9910}, 030 (1999)
  [arXiv:hep-th/9909101].
  %%CITATION = JHEPA,9910,030;%%
  }
  
  \lref\rharmarkoberst{
  T.~Harmark and N.~A.~Obers,
  ``Thermodynamics of the near-extremal NS5-brane,''
  Nucl.\ Phys.\  B {\bf 742}, 41 (2006)
  [arXiv:hep-th/0510098]. 
  T.~Harmark and N.~A.~Obers,
  ``New phases of thermal SYM and LST from Kaluza-Klein black holes,''
  Fortsch.\ Phys.\  {\bf 53}, 536 (2005)
  [arXiv:hep-th/0503021].
  T.~Harmark, V.~Niarchos and N.~A.~Obers,
  ``Instabilities of near-extremal smeared branes and the correlated  stability
  conjecture,''
  JHEP {\bf 0510}, 045 (2005)
  [arXiv:hep-th/0509011].}

 \lref\rharmarkobers{
  T.~Harmark and N.~A.~Obers,
  ``Hagedorn behavior of little string theories,''
  arXiv:hep-th/0010169.
  
  T.~Harmark and N.~A.~Obers,
  ``Hagedorn behaviour of little string theory from string corrections to
  NS5-branes,''
  Phys.\ Lett.\  B {\bf 485}, 285 (2000)
  [arXiv:hep-th/0005021].
   }
  
   \lref\rmaldastro{
  J.~M.~Maldacena and A.~Strominger,
  ``Semiclassical decay of near-extremal fivebranes,''
  JHEP {\bf 9712}, 008 (1997)
  [arXiv:hep-th/9710014].
  %%CITATION = JHEPA,9712,008;%%
  }
  
  \lref\rchs{
   C.~G.~.~Callan, J.~A.~Harvey and A.~Strominger,
  ``Supersymmetric string solitons,''
  arXiv:hep-th/9112030.
  %%CITATION = HEP-TH/9112030;%%
  }
  
  \lref\rdorey{
   N.~Dorey,
  ``A new deconstruction of little string theory,''
  JHEP {\bf 0407}, 016 (2004)
  [arXiv:hep-th/0406104].
  %%CITATION = JHEPA,0407,016;%%
  }
  
  \lref\rmotl{
  N.~Arkani-Hamed, A.~G.~Cohen, D.~B.~Kaplan, A.~Karch and L.~Motl,
  ``Deconstructing (2,0) and little string theories,''
  JHEP {\bf 0301}, 083 (2003)
  [arXiv:hep-th/0110146].
  %%CITATION = JHEPA,0301,083;%%
  }

  \lref\radshag{J.~L.~F.~Barbon and E.~Rabinovici,
  ``Closed-string tachyons and the Hagedorn transition in AdS space,''
  JHEP {\bf 0203}, 057 (2002)
  [arXiv:hep-th/0112173].
  %%CITATION = JHEPA,0203,057;%%
  }
  
  \lref\ncen{ J.~L.~F.~Barbon and E.~Rabinovici,
  ``On 1/N corrections to the entropy of noncommutative Yang-Mills  theories,''
  JHEP {\bf 9912}, 017 (1999)
  [arXiv:hep-th/9910019].
  %%CITATION = JHEPA,9912,017;%%
  }
  
  \lref\rthresholds{J.~L.~F.~Barbon, I.~I.~Kogan and E.~Rabinovici,
  ``On stringy thresholds in SYM/AdS thermodynamics,''
  Nucl.\ Phys.\  B {\bf 544}, 104 (1999)
  [arXiv:hep-th/9809033].
  %%CITATION = NUPHA,B544,104;%%
  }
  
  \lref\rwkbs{ J.~L.~F.~Barbon,
  ``Remarks on thermal strings outside black holes,''
  Phys.\ Lett.\  B {\bf 339}, 41 (1994)
  [arXiv:hep-th/9406209].
  J.~L.~F.~Barbon,
  ``Horizon divergences of fields and strings in black hole backgrounds,''
  Phys.\ Rev.\  D {\bf 50}, 2712 (1994)
  [arXiv:hep-th/9402004].
  %%CITATION = PHRVA,D50,2712;%%
  }
   \lref\rrangamani{M.~Rangamani,
  ``Little string thermodynamics,''
  JHEP {\bf 0106}, 042 (2001)
  [arXiv:hep-th/0104125].
 }
 
 \lref\rreall{
 H.~S.~Reall,
  ``Classical and thermodynamic stability of black branes,''
  Phys.\ Rev.\  D {\bf 64}, 044005 (2001)
  [arXiv:hep-th/0104071].
 }
 
 \lref\rglaflamme{
 R.~Gregory and R.~Laflamme,
  ``Black strings and p-branes are unstable,''
  Phys.\ Rev.\ Lett.\  {\bf 70}, 2837 (1993)
  [arXiv:hep-th/9301052].
 }
 
 \lref\rmarolf{
 D.~Marolf,
  ``Asymptotic flatness, little string theory, and holography,''
  JHEP {\bf 0703}, 122 (2007)
  [arXiv:hep-th/0612012].
  D.~Marolf and A.~Virmani,
  ``Holographic Renormalization of Gravity in Little String Theory Duals,''
  [arXiv:hep-th/0703251].
  }
  
  \lref\rcigar{
   S.~Elitzur, A.~Forge and E.~Rabinovici,
  ``Some global aspects of string compactifications,''
  Nucl.\ Phys.\  B {\bf 359}, 581 (1991).
  Mandal, A.~M.~Sengupta and S.~R.~Wadia,
  ``Classical Solutions Of Two-Dimensional String Theory,''
  Mod.\ Phys.\ Lett.\  A {\bf 6}, 1685 (1991).
 E.~Witten,
  ``On string theory and black holes,''
  Phys.\ Rev.\  D {\bf 44}, 314 (1991).
  }
  
  \lref\ratickw{
  J.~J.~Atick and E.~Witten,
  ``The Hagedorn Transition and the Number of Degrees of Freedom of String
  Theory,''
  Nucl.\ Phys.\  B {\bf 310}, 291 (1988).
  %%CITATION = NUPHA,B310,291;%%
  }
  
  \lref\rgpy{
  D.~J.~Gross, M.~J.~Perry and L.~G.~Yaffe,
  ``Instability Of Flat Space At Finite Temperature,''
  Phys.\ Rev.\  D {\bf 25}, 330 (1982).
  %%CITATION = PHRVA,D25,330;%%
  }
  
  \lref\rgiboptical{
   G.~W.~Gibbons and M.~J.~Perry,
  ``Black Holes And Thermal Green's Functions,''
  Proc.\ Roy.\ Soc.\ Lond.\  A {\bf 358}, 467 (1978).
  %%CITATION = PRSLA,A358,467;%%
  }
  
  \lref\rhpage{
   S.~W.~Hawking and D.~N.~Page,
  ``Thermodynamics Of Black Holes In Anti-De Sitter Space,''
  Commun.\ Math.\ Phys.\  {\bf 87}, 577 (1983).
  }
  
  \lref\radscon{
  E.~Witten,
  ``Anti-de Sitter space, thermal phase transition, and confinement in  gauge
  theories,''
  Adv.\ Theor.\ Math.\ Phys.\  {\bf 2}, 505 (1998)
  [arXiv:hep-th/9803131].
  }
  
  \lref\roptical{
   J.~S.~Dowker and G.~Kennedy,
  ``Finite Temperature And Boundary Effects In Static Space-Times,''
  J.\ Phys.\ A  {\bf 11}, 895 (1978).
  %%CITATION = JPAGB,A11,895;%%
  }
  
  \lref\rgiveonkrs{
  A.~Giveon, D.~Kutasov, E.~Rabinovici and A.~Sever,
  ``Phases of quantum gravity in AdS(3) and linear dilaton backgrounds,''
  Nucl.\ Phys.\  B {\bf 719}, 3 (2005)
  [arXiv:hep-th/0503121].}
  
  \lref\rberkleb{
   M.~Bershadsky and I.~R.~Klebanov,
  ``Genus one path integral in two-dimensional quantum gravity,''
  Phys.\ Rev.\ Lett.\  {\bf 65}, 3088 (1990).
  %%CITATION = PRLTA,65,3088;%%
  }
  
  \lref\rjabbari{
  D.~Sadri and M.~M.~Sheikh-Jabbari,
  ``Integrable spin chains on the conformal moose,''
  JHEP {\bf 0603}, 024 (2006)
  [arXiv:hep-th/0510189].
  }
  
  \lref\rozalisha{
  M.~Alishahiha, Y.~Oz and M.~M.~Sheikh-Jabbari,
  ``Supergravity and large N noncommutative field theories,''
  JHEP {\bf 9911}, 007 (1999)
  [arXiv:hep-th/9909215].
  %%CITATION = JHEPA,9911,007;%%
}
\lref\rmaldarusso{
J.~M.~Maldacena and J.~G.~Russo,
  ``Large N limit of non-commutative gauge theories,''
  JHEP {\bf 9909}, 025 (1999)
  [arXiv:hep-th/9908134].}
  
  \lref\ritzhakihashimoto{
  A.~Hashimoto and N.~Itzhaki,
  ``Non-commutative Yang-Mills and the AdS/CFT correspondence,''
  Phys.\ Lett.\  B {\bf 465}, 142 (1999)
  [arXiv:hep-th/9907166].}
  
  \lref\rdecons{
  N.~Arkani-Hamed, A.~G.~Cohen and H.~Georgi,
  ``(De)constructing dimensions,''
  Phys.\ Rev.\ Lett.\  {\bf 86}, 4757 (2001)
  [arXiv:hep-th/0104005].}
  
  \lref\radamsfab{
  A.~Adams and M.~Fabinger,
  ``Deconstructing noncommutativity with a giant fuzzy moose,''
  JHEP {\bf 0204}, 006 (2002)
  [arXiv:hep-th/0111079].}
  
  \lref\rdoreyp{
  N.~Dorey,
  ``S-duality, deconstruction and confinement for a marginal deformation of  N
  %= 4 SUSY Yang-Mills,''
  JHEP {\bf 0408}, 043 (2004)
  [arXiv:hep-th/0310117].
  }
  
  \lref\rmyers{
  R.~C.~Myers,
  ``Dielectric-branes,''
  JHEP {\bf 9912}, 022 (1999)
  [arXiv:hep-th/9910053].}
  
  \lref\rus{
   J.~L.~F.~Barbon and E.~Rabinovici,
  ``Extensivity versus holography in anti-de Sitter spaces,''
  Nucl.\ Phys.\  B {\bf 545}, 371 (1999)
  [arXiv:hep-th/9805143].
  %%CITATION = NUPHA,B545,371;%%
  }

%%%%%%%%TEXT%%%%%%%%%%%%%%%%%%%%%%%%%%%%%%%%%%%%%%%%%%%%%%%%%%%%%%%%%
%%%%%%%%%%%%%%%%%%%%%%%%%%%%%%%%%%%%%%%%%%%%%%%%%%%%%%%%%%%%%%%%%%%%%
%%%%%%%%%%%%%%%%%%          title page       %%%%%%%%%%%%%%%%%%%%%%%%%
%%%%%%%%%%%%%%%%%%%%%%%%%%%%%%%%%%%%%%%%%%%%%%%%%%%%%%%%%%%%%%%%%%%%%%

\baselineskip=15pt

\line{\hfill IFT UAM/CSIC-2007-37}

\vskip 0.7cm

\Title{\vbox{\baselineskip 12pt\hbox{}
 }}
{\vbox {\centerline{Deconstructing the  }
\vskip10pt
\centerline{ Little Hagedorn Holography}
}}

\vskip 0.5cm

\centerline{$\quad$ {\caps
Jos\'e L.F. Barb\'on$^\dagger$,
 Carlos A. Fuertes$^\dagger$
 and
Eliezer Rabinovici$^\star$
}}
\vskip0.7cm

\centerline{{\sl  $^\dagger$ Instituto de F\'{\i}sica Te\'orica IFT UAM/CSIC }}
\centerline{{\sl  C-XVI,
 UAM, Cantoblanco 28049. Madrid, Spain }}
\centerline{{\tt jose.barbon@uam.es, carlos.fuertes@uam.es}}

\vskip0.2cm

\centerline{{\sl $^\star$
Racah Institute of Physics, The Hebrew University }}
\centerline{{\sl Jerusalem 91904, Israel}}
\centerline{{\tt eliezer@vms.huji.ac.il}}

\vskip0.7cm

\centerline{\bf ABSTRACT}

 \vskip 0.3cm

 \noindent

We study aspects of the  thermodynamics of Little String Theory, using its geometrical  definition  in critical ten-dimensional string theory. We find that bulk radiation effects tend to  screen the Hagedorn behaviour of the theory, in contrast to the behaviour in the  AdS system background.
The resulting density of states of the system, when stable, is described by a seven-dimensional nonrelativistic gas. This requires modifications of the holographic Little Hagedorn picture. Using deconstructions we suggest such
modifications. The model is embedded into a system which has an ultraviolet fixed point with an AdS description.   We investigate the thermodynamical  properties of these UV completed  models.  It is found that the Hagedorn regime survives  in a finite band of  superheated states that eventually   decay into the plasma phase of the conformal field theory that serves as UV regulator.  This is manifested in a first-order phase transition that is driven by  radiative corrections. 

\vskip 1cm

\Date{June 2007}

\vfill

\vskip 0.1cm

%%%%%%%%%%%%%%%%%%%%%%%%%%%%%%%%%%%%%%%%%%%%%%%%%%%%%%%%%%%%%%%%%%%%%%

%\draft

%%%%%%%%%%%%%%%%%%%%%%%%%%%%%%%%%%%%%%%%%%%%%%%%%%%%%%%%%%%%%%%%%%%%%%%%%%
%%%%%%%%%%%%                text begins                        %%%%%%%%%%%
%%%%%%%%%%%%%%%%%%%%%%%%%%%%%%%%%%%%%%%%%%%%%%%%%%%%%%%%%%%%%%%%%%%%%%%%%%

\baselineskip=15pt

\newsec{Introduction}

\noindent

Asymptotically free and conformal local field theories seem to be self-contained at all energy scales and in particular do not need an ultraviolet (UV) completion. In the presence of gravity such a completion is required and string theory stands out as the main  candidate. The advent of holography in the AdS/CFT context has softened the difference between these underlying structures and it was discovered that some string theories containing gravity are actually equivalent to well defined field theories which do not contain gravity explicitly \refs\radscft.  In the course of constructing pairs of duals a new type of underlying structure was suggested to exist, the Little String Theories (LST) \refs{\rlstbasic, \rdlcq, \raharonyrew}. These models are   supposed to have a Hagedorn density of states at very high energies, 
\eqn\hag{
  \Omega(E)\approx  {\exp\left( \beta_H \,E\right)
  \over E^{1+\gamma}}\;,}
which is characteristic of weakly-coupled string theories. The fundamental length scale $\beta_H$ is the inverse Hagedorn temperature, $T_H^{-1}$,  and the exponent $\gamma$ controls the thermodynamical stability (see \refs\rtouring\ for a review and a collection of references). Although LST models do not contain gravity they   do posses  T-duality, which is not a possible property of a local field theory. LST is thus a system which is intermediate in this sense between string theory and field theory, perhaps strongly related to the large-$N$ limit of QCD.  Many of these properties can be  inferred from the holographic description of the theory in terms of a linear-dilaton background in type II string theory \refs\rfivebraneholo. A fundamental feature of the bulk description is the fact that the Hagedorn spectrum \hag\ is geometrically realized as a black-fivebrane spectrum \refs\rmaldafiveb. 

The Hagedorn spectrum is indeed a distinctive property of free string theory. In the strong interactions such a spectrum and the presence of the limiting temperature which it entails led to the suggestion that there is a different underlying structure to the hadrons in QCD. The constituents are in this case the quarks and gluons. For any finite amount of colors, $N$, there exists a large enough energy for which the density of states is that of the QCD field theory. In terms of temperature there is no limiting temperature. This is difficult to discover when  the
 theory is truly free (infinite $N$) but gets exposed for finite $N$ when the system has interactions. The quest for a possible underlying structure of strings makes it worthwhile to scrutinize all systems which seem to have a Hagedorn density of states. One clue in this quest is provided by the  marginal stability of these systems, this in the sense that the microcanonical temperature function,
 \eqn\microcant{
 T(E) = \Omega(E) \,\left({\pt \Omega (E) \over \pt E} \right)^{-1}
 }
   starts up being a constant temperature denoted by $T_H = \beta_H^{-1}$. A small correction that turns $T(E)$ into an increasing function of the energy, $E$,  provides stability to the system through its positive specific heat. On the other hand a correction leading to  the  opposite behaviour endows the system with a negative specific heat, signalling a thermodynamical instability. It could also be an indication that the theory undergoes a first-order transition to a stable high-energy phase, if such a phase exists. The sources of these modifications can be traced to features of the background such as curvature and finite-size effects as well as
  to perturbative interaction corrections and nonperturbative effects, affecting the precise value of the coefficient $\gamma$ in \hag, as well as introducing other types of functional dependencies.
  
The main example is the case of the interacting critical, ten dimensional string gas.
At energy densities of order $g_s^{-2}$ in string units an underlying structure emerges: the nucleation of small black holes induces their negative specific heat, making the system unstable to the unlimited growth of these black holes. The system can be stabilized by an infrared (IR) regularization; an AdS box where  the small black holes end up growing to the size of the large AdS black hole at the Hagedorn temperature of the bulk, i.e. $T_H = \lambda^{1/4} T_{HP}$, where $T_{HP}$ is the Hawking--Page temperature of the AdS space \refs{\rhpage, \radscon},  and $\lambda$ is the 't Hooft coupling of the dual CFT. Hence, the ten-dimensional Hagedorn phase appears as a superheated unstable state of the full CFT, that decays into the stable plasma phase \refs\radshag. 
In this paper, we shall argue that LSTs present a conceptually  similar picture, albeit slightly modified by  the specific subtleties of the LST holographic map. 
 
   LSTs can be  canonically constructed as the decoupling limit of the type
 IIA string dynamics on a stack of  $N$ coincident NS5-branes, with vanising string coupling,  $g_s \rightarrow 0$, and fixed  string length $\ell_s = m_s^{-1}$. 
 The dual geometry is described by adding to the world-volume coordinates $(t, y_5) \in {\bf R}^{1+5}$  a  `tube' with a holographic `radial' coordinate $z\in {\bf R}$ and a transverse three-sphere of radius $R= \sqrt{\alpha' N} \equiv \ell_s \sqrt{N}$, threaded by $N$ units of Neveu--Schwarz flux. In addition, there is a linear dilaton profile whose slope is
 $1/R$:
 \eqn\hololst{
 ds^2 = -dt^2 + dy_5^2 + dz^2 + R^2 \,d\Omega_3^2\;, \qquad e^\phi = e^{-z/R}\;.
 }
 This is the decoupling limit which retains  stringy features,
  such as T-duality and the Hagedorn density of states with effective inverse temperature
   $\beta_H = T_H^{-1} = 2\pi \ell_s \sqrt{N}$. In this paper, we deal mostly with the situation in which
   $N$ is parametrically large but {\it finite}. A crucial aspect of this large $N$ regime is the fact that the Hagedorn spectrum is {\it not} related to gas of long strings in the geometry \hololst\ but to the black-brane states with asymptotics given by \hololst. 
   This model can be defined with either type IIA or IIB fivebranes, as they are related by a T-duality, but the low-energy sector of these two models is quite different. The type IIA LST flows in the infrared to the six-dimensional  $(2,0)$ conformal field theory (CFT), whereas the type IIB counterpart has a six-dimensional Yang--Mills theory with $(1,1)$ supersymmetry. 
   
   One of the striking features of \hololst\ is the ocurrence of a {\it flat} seven-dimensional Minkowski metric by combining the world-volume coordinates with the   radial `holographic coordinate' $z$. This is rather surprising for a system that should presumably describe a six-dimensional theory. Unlike the more familiar AdS case, there are no gravitational red-shift effects in the $z$ direction that effectively confine normalizable modes within a finite range of $z$. This feature has been related to the nonlocality of the dual LST theory (c.f. \refs{\rpeetpol, \rfivebraneholo, \rlszlst}) and presents LST as an interesting laboratory for ideas about holography in asymptotically flat spacetimes (c.f. \refs\rmarolf\ for a recent discussion).    In fact, the linear dilaton slope is responsible for a `partial containment' of excitations in the radial direction, inducing a {\it finite} mass gap
   $M_{\rm eff} = 1/R$ for propagation along the $z$ coordinate \refs\rminwallasei. The finite height of this barrier means that a thermal gas of particles at the LST Hagedorn temperature $T_H = (2\pi R)^{-1}$, contributing to the one-loop corrections of thermodynamic functions, 
   is effectively nonrelativistic when propagating down the essentially  infinite `tube'. Although subject to a Boltzman suppression, the corresponding thermodynamical functions behave extensively in $z$. There is an induced infrared divergence, 
   thus undoing  the loop expansion of the LST, at least to the extent that one uses the geometry \hololst\ to do computations. One can proceed by imposing an {\it ad hoc}  cutoff at large values of $z$ in the linear-dilaton background, but then it should be checked that the results, and the order of the various limits, are independent of the cutoff, or else a physical interpretation of the cutoff must be provided.
 We find that the discussion of the thermodynamic stability of LST theories is intimately connected with the fate of the volume divergence in the holographic direction. This is done by first uncovering irregularities trying to set up  a canonical framework and then uncovering the sources of these irregularities in a microcanonical analysis. Indeed it turns out that the bulk properties in the $z$ direction can dominate the properties of the whole system leading to a less familiar nonrelativistic but extensive gas of strings. We derive this and discuss the effects of such a gas in some detail. In particular this gas turns out not to have a Hagedorn density of states.

 In order to be able to better disentangle the various limits involved we find it useful to  embed the LST in a  CFT at high energies. In this way one is guaranteed a standard implementation of holography in the deep ultraviolet regime, as well as a physical interpretation for whatever instabilities are found. The strategy is analogous to  the one followed in \refs{\radshag, \rbek}, in which the Hagedorn behaviour of ten-dimensional critical IIB strings was given a finite-volume regularization in the standard ${\rm AdS}_5 \times {\bf S}^5$ background. The immediate payback of such a procedure is that one knows the stable state at high temperature for this system, namely a large ${\rm AdS}_{d+1}$  black hole with temperature $T(E) \sim
 E^{1/d}$, for a CFT fixed point with $d$ spacetime dimensions. The results are of course not universal but depend on the specific UV completion. However for any such completion the transient role of the Hagedorn behavior is clearly uncovered.
The particular embedding of the LST into a UV fixed point will use the ideas of `deconstruction' \refs\rdecons. Namely, we will consider a lattice approximation to the LST, where the lattice is engineered on the Higgs branch of a lower-dimensional field theory with a more or less standard UV fixed point \refs{\rmotl, \rdorey}. This has been done already in the literature for other purposes for type IIB, the treatment for type IIA presented here is novel.

 Our results indicate that the fate of the Hagedorn transition in the so-regularized LST theories is ultimately very similar to that of standard critical strings in AdS spaces. Namely we find a first-order phase transition that `censors' the Hagedorn regime as in \refs\rthresholds. The Hagedorn density of states pertains yet again to a band of superheated states that are either unstable or metastable, depending on the value of the energy.  There is however some difference, since here the first-order transition does not occur at the classical level but is driven by the quantum mechanical radiative corrections. The LST is found not to clearly decouple the boundary and the bulk components. One may try to argue that by following the dependence on the volume cutoff
one could isolate a non leading defect like the behaviour of just the horizon of the blackbrane. It is yet to be
demonstrated how to actually implement such a program in a unique manner.

 The paper is organized as follows. In Section 2 we review some well-known aspects of LST thermodynamics, emphasizing the issue of string loop corrections. In Section 3 we discuss the
 extensivity of radiation in the asymptotic tube geometry. In Section 4 we elaborate on the impact of such extensivity in the holographic properties of the LST background and obtain a non relativistic string gas. In Section 5 we contrast the AdS and LST holographic properties. We focus on the breakdown of holography in some cases by using an effective potential picture for various D$p$ branes, $p=5$ being  a border case. In Section 6 we discuss the off-shell thermodynamical stability of LST models with an explicit regularization of the tube length, this is done by both a canonical and microcanonical analysis. 
 The physical role of the radial IR cutoff is discussed. In Section 7 we present  several deconstructed models and their corresponding UV completions. In Section 8 we study the thermodynamics of the deconstructed models.

\newsec{Review of LST thermodynamics}

\noindent

According to the LST holographic dictionary, the classical approximation to the thermal ensemble is described geometrically by
the decoupling limit of the near-extremal NS5-brane metric:
\eqn\blackns{
ds^2 = -dt^2 \left(1-{r_0^2 \over r^2}\right) + dy_5^2 + {R^2 dr^2 \over r^2 -r_0^2} + R^2 \,d\Omega_3^2\;,
}
and dilaton $e^{\phi} = g_s R/r$, where we have introduced a spherical radial coordinate $r=g_s R \exp(z/R)$. The associated Hawking temperature is independent of
the horizon radius $r_0$, and given by $T(r_0) = (2\pi R)^{-1}$. The energy, defined with reference to that of an extremal NS5-brane embedded in ten-dimensional Minkowski space, is
\eqn\energ{
E (r_0)= {V_5 m_s^8 r_0^2 \over (2\pi)^5 g_s^2}\;,
}
whereas the Bekenstein--Hawking entropy satisfies the Hagedorn law $S= \beta_H E = 2\pi R E$. As a result, the classical approximation to the Euclidean action of \blackns, or free energy, vanishes for all values of $r_0$:
\eqn\fenn{
I(r_0) = \beta_H \,E(r_0) - S(r_0)=0
\;.}
  This strict Hagedorn behaviour is a property of the decoupling limit that isolates the
LST dual metric \blackns. Retaining the full, asymptotically flat, ten-dimensional CHS background \refs\rchs,
\eqn\fullten{
ds^2 = -dt^2 \left(1-{r_0^2 \over r^2}\right) + dy_5^2  + \left(1+ {R^2 \over r^2}\right) \left({dr^2 \over 1-{r_0^2 /r^2}}  + r^2 \,d\Omega_3^2 \right)\;,\qquad e^{2\phi} = g_s^2 \left(1+{R^2 \over r^2} \right)
}
one finds a temperature function 
$$
T(r_0) = {1\over 2\pi R\sqrt{1+r_0^2 / R^2}}
$$
that, via Eq. \energ,   decreases monotonically as a function of the energy, resulting in a system of negative specfic heat.

We can now use the location of the event horizon, $r_0$, as a probe of the local propeties of the background. 
The  local value of the string coupling at the horizon grows from $O(g_s)$ at the top of the tube, to $O(1)$ at
$r_0 \sim r_s = g_s R$. At lower radii we must use some S-dual description, with details depending on  whether we work in type IIA or type IIB (c.f. \refs\ritzhaki).

In type IIA the S-dual  is an eleven-dimensional metric of $N$ M5-branes smeared over the eleven-dimensional circle of length $R_{11} = g_s \ell_s$. In eleven-dimensional units, this S-duality matching takes place at $r_0\sim r_s = R_{11} \sqrt{N}$. The thermodynamics of a near-extremal solution with horizon radii $r_0 < r_s$   continues to be Hagedorn-like until we reach radius 
around $r_0 \sim r_{ H} = R_{11} = g_s R /\sqrt{N}$, where  the smeared M5-branes localize on the circle, and we flow to
 the ${\rm AdS}_7 \times {\bf S}^4$
 background
that characterizes the dual $(2,0)$ CFT in six dimensions. \foot{The details of this localization transition are interesting, involving nontrivial transients as explained in \refs\rharmarkoberst.} The standard thermodynamics of ${\rm AdS}_7$ black branes yields  entropy $S_{\rm CFT} \sim N^3 V_5 T^5$ and energy $E_{\rm CFT}
\sim N^3 V_5 T^6$  for the rank-$N$ $(2,0)$ CFT. The matching to
the Hagedorn regime  $S_{\rm Hag} =E/T_H$ occurs  at radii $r_0 \sim r_{ H} $. At this  point the thermodynamics of the $(2,0)$ CFT and the Hagedorn-like black brane match in temperature, $T=T_H$,  energy $E_H \sim N^3 V_5 T_H^6$ and
entropy $S_H \sim N^3 V_5 T_H^5$ (notice the peculiar scaling of the $(2,0)$ CFT entropy, corresponding to $O(N^3)$ degrees of freedom).

In type IIB theory, the S-dual transition at $r_0\sim r_s = R g_s$ yields the near-extremal D5-brane metric, which becomes highly
curved at $r_0 \sim r_{ H} \sim R g_s /N$. At this point we match to a non-geometrical description in terms  of  six-dimensional SYM with
rank $N$. The corresponding thermodynamical functions  $S_{\rm YM} \sim N^2 V_5 T^5$ and $E_{\rm YM} \sim N^2 V_5 T^6$ (in the free approximation) match  the `Hagedorn entropy and energy', $S_H \sim N^2 V_5 T_H^5$, and $E_H \sim N^2 V_5 T_H^6$ at the threshold $T=T_H$.

\fig{\sl   The different regions of the bulk type IIA background. The local string coupling grows towards smaller radii in the NS5 `tube', becoming of $O(1)$ at $r\sim r_s = g_s R$. At lower radii the system is well approximated by the uplifted solution to eleven dimensions, i.e.  the smeared ${\widetilde {\rm M5}}$-brane solution, which localizes below $r\sim r_H = g_s R /\sqrt{N}$ and flows to the ${\rm AdS}_7 \times {\bf S}^4$ dual of the $(2,0)$ CFT in six dimensions. In the type IIB case, the ${\widetilde {\rm M5}}$ phase
is replaced by the near-horizon D5-brane background, and the matching at $r\sim r_H$ takes the system to a non-geometrical phase described by weakly-coupled Yang--Mills theory. }{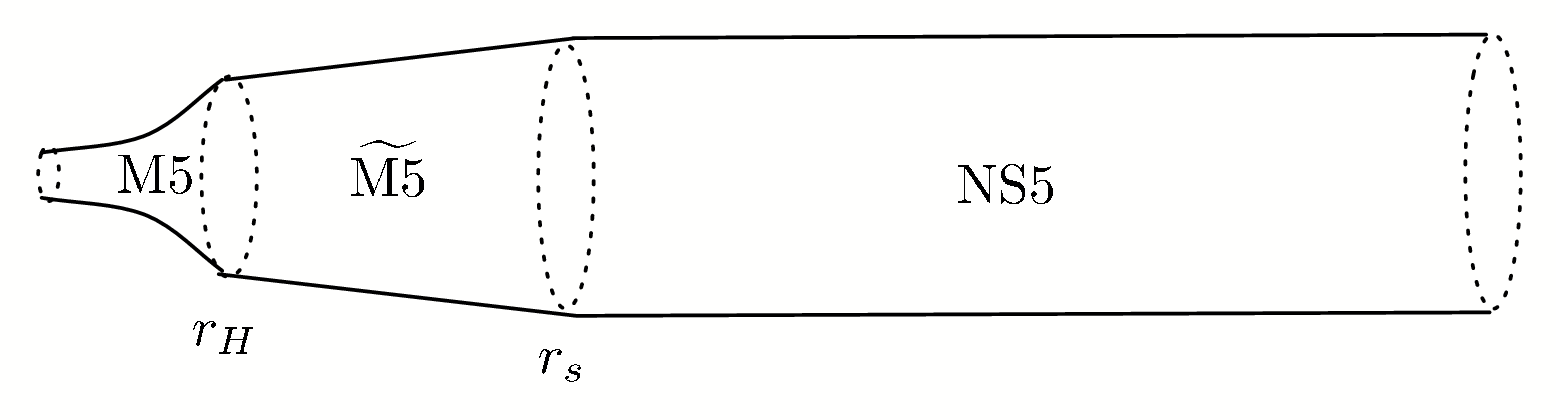}{4truein}

To summarize, the leading large-$N$ thermodynamics of the LST system can be captured by the Hagedorn-like black-brane entropy in the whole range $r_{\rm H} < r_0 < \infty$. \foot{There is a narrow   region around the S-duality threshold,  $r_0\sim r_s =g_s R$, where loop corrections are of $O(1)$, but the classical thermodynamic functions are insensitive to this transition.}
At the lower limit, $r_0\sim r_{ H} $, the system matches to weakly coupled Yang--Mills in the type IIB case, which does not admit a geometrical description. In the type IIA case, however, the low energy $(2,0)$
CFT does admit a geometrical description at large $N$, in terms of the ${\rm AdS}_7 \times {\bf S}^4$
background of eleven-dimensional supergravity. In this case the transition at $r_0 \sim r_{ H}$ is a matching between the AdS region at $r<r_{ H}$ and the `tube' region of the metric at $r_{ H}<r < \infty$. The type IIA `tube' is eleven-dimensional in the region $r_{ H} < r<g_s R$, and ten-dimensional for $r >g_s R$. Strictly speaking, the  geometry is only `tube-like' in the ten-dimensional region, but we refer to whole geometry above the AdS matching as `the tube', because the classical thermodynamic functions do not notice the strong-coupling transition at $r_0\sim g_s R$.

\subsec{Loop corrections and stability}

\noindent

A central question regarding the deep LST regime, at energies $E\gg E_H$, is whether the `Hagedorn plateau', i.e. the constant function  $T(E) = T_H$, is corrected  upwards or downwards. The choice among these alternatives would determine  the thermodynamical  stability of this phase. Equivalently, we seek to lift the degeneracy of the Euclidean action as a function of the horizon location
\fenn.

  We would like to answer this question in the fully decoupled background \blackns. Of course, different embeddings or deformations of the decoupled metric might have particular stability properties. For example, keeping the undecoupled system, as in \fullten, results in an unstable phase.  Ideally, we would like to disentangle whatever instabilities could be induced by the particular UV completion of the LST system, from the intrinsic instabilities of the self-contained string theory defined in \blackns.  In this section, we will identify the radiation in the `tube' region as the main contender in deciding the fate of the function $T(E)$ for this system. 
  
    The corrections to the function $T(E)$ come in principle from both $\alpha'$ and $g_s$ loop effects in string theory. Short-distance $\alpha'$ corrections at string tree-level  do not seem capable of removing the Hagedorn degeneracy. The Euclidean  background contains the coset $SL(2, {\bf R})/U(1)$ as a factor  \refs\rcigar,  an exact worldsheet CFT description (at least in the type IIA case), for any value of the horizon location, and with a fixed value of the asymptotic temperature. Hence, no $\alpha'$ corrections are expected for the thermodynamical functions at string tree level.

An estimate of the one string-loop
correction to the free energy was presented in \refs\rkut, where only
radiation far from the horizon was taken into account. In fact, for $N\gg 1$, the calculation can be performed reliably in the low-energy supergravity approximation. The effect of higher string modes, at energies of $O(m_s)$, can be incorporated by adding higher-dimensional operators to the effective supergravity model, whereas the effect of the light modes  may be dealt with directly.

A contribution to the free energy can be split into the part which is extensive in the length of the tube,
with effective geometry ${\bf R}^{1+6} \times {\bf S}^3$ and a linear dilaton profile, and the part coming from the horizon region,
with curvature of $O(1/R^2)$.
In general, loop corrections at the horizon location are of relative order
$$
e^{2\phi(r_0)} = {g_s^2 R^2 \over r_0^2}  =
N^{4-\alpha} \;{E_H \over E}\;,
$$
where $\alpha =3$ in the type IIA theory, and $\alpha=2$ in the type IIB theory.  Hence, the corrections to the microcanonical temperature function coming only from local effects near the horizon are
\eqn\col{
{\delta T(E) \over T_H}\Big|_{\rm horizon} \approx
 C{N^{4-\alpha} E_H \over E}\;,
}
with $C$ an unknown coefficient of $O(1)$ (c.f. \refs{\rharmarkobers, \rcoiled}). This expression is valid for $E$ large enough so that the correction term is small. At $E\sim N^{4-\alpha} E_H$ (equivalently $r_0 \sim Rg_s$), the loop corrections are very significant, and at lower energies we must use the dual description. For example, in the type IIA model, the eleven-dimensional supergravity background has as quantum expansion parameter $\ell_p^2 / \CR^2$, where $\ell_p$ is the eleven-dimensional Planck length and $\CR$ is a characteristic radius of curvature. At the LST threshold, $r_0\sim R_{11}$, we have $\CR\sim
\ell_p \,N^{1/3}$, so that quantum corrections to $T(E) /T_H$ are of relative order $O(N^{-2/3})$.  Thus, the supergravity approximation is uniformly accurate for large $N$, except in the immediate vicinity of the strong-coupling threshold $r_0 \sim g_s R$.

 Conversely, in the type IIB model the string loop corrections are of $O(1)$ at  $r_0\sim g_s R$. At lower radii,  we  switch to the S-dual near-extremal D5-brane
metric. In this case, the local string coupling decreases towards small radius, but the $\alpha'$ corrections become dominant, producing $O(1)$ corrections to \col\ at the threshold to the low-energy
Yang--Mills description.  Hence, the type IIB model has small  loop corrections localized at the horizon provided $N$ is sufficiently large, except for the vicinity of the LST threshold, $r_0 \sim r_H$ and the
S-duality threshold $r_0 \sim r_s$. 

The remaining contributions to the one-loop corrections can be associated to the effect of string modes propagating `in the tube'. These  are extensive in the  length of the tube (the extension of the $z$ coordinate) and
therefore dominate the horizon effects for a sufficiently long tube.
The contribution to the free energy of this string gas with extensive behaviour in the $z$ coordinate was calculated in ref. \refs\rkut, with the result
\eqn\radfe{
\beta_H \Delta F_{\rm tube} \equiv \Delta I_{\rm tube} \approx -C_I\,V_5 \,\Delta z\, T_H^{\,6}\;,
}
where $C_I$ is a positive numerical constant of $O(1)$ and $\Delta z$ is the length of the tube containing the string gas at temperature $T_H$. Notice that we can obtain the form of this expression
by simple dimensional analysis and the asumption of extensivity in $V_5 \Delta z$, since the LST
has only one intrinsic energy scale, $T_H \sim 1/R$, in the limit $\ell_s T_H \sim N^{-1/2} \ll 1$. In this limit, the degrees of freedom are well accounted for by  the low-energy supergravity approximation, and  the same dimensional argument
gives an entropy with a seven-dimensional, field-theoretical scaling:
\eqn\rades{
S_{\rm tube}  \approx C_S \,V_5 \,\Delta z \,T_H^{\,6}
\;,}
with $C_S$ another positive constant of $O(1)$.

 If the decoupled metric
 \blackns\ is taken at face value, we have   $\Delta z =\infty$ and the radiation appears to dominate the
 canonical ensemble.  Thus, we face a situation very similar to that of a canonical ensemble in flat-space gravity, which does not admit a rigorous definition because of the Jeans instability (c.f. for example \refs\rgpy). In the LST case, the situation is not quite as ill-defined as in standard flat Minkowski space, because the local string coupling (and thus the local value of Newton's constant) decreases exponentially at large $z$. This means that the radiation system  in the tube is asymptotically free at large $z$, and not particularly prone to the Jeans instability. On the other hand,  the free radiation does contribute a large density of states, spoiling the (six-dimensional) holographic scaling of the entropy.  This apparent clash between
radiation thermodynamics and holography will be the subject of the Sections 3 and 4.

 If we impose an {\it ad hoc} regularization of the tube length, $\Delta z = z_\Lambda - z_0 $, we can write $\Delta z = \shalf R \log(E_\Lambda / E_b)$, where $E_b$ is the energy stored in the black-brane horizon at the classical level, Eq.  \energ, and $E_\Lambda$ is the classical energy of a black brane with horizon radius $z_0 = z_\Lambda$.  Then, adding \rades\ to the classical horizon entropy $\beta_H E_b$, we find a one-loop corrected entropy function 
 \eqn\entkut{
 S(E_b) = \beta_H \,E_b - (1+\gamma)\,\log(E_b /E_\Lambda)\;, \qquad 1+\gamma= \shalf C_S \,R\,V_5\, T_H^{\,6}\;.
 }
 This parametrization was used in  \refs\rkut\  to argue  that the LST thermodynamics is unstable. Notice, however, that the energy $E_b$ appearing in \entkut\ includes only the classical black-brane energy. A complete analysis of the stability of the system requires that we compute the function $S(E)$, with $E$ the {\it total} energy in black brane {\it and} radiation in an equilibrium configuration.  This suggests that the qualitative physics implied by \entkut\ will be valid only in situations where the total contribution of radiation to the energy is subdominant. At any rate, \entkut\ gives an estimate of the critical exponent $\gamma$ appearing in \hag. A most peculiar feature,  when compared to  similar exponents in the theory of critical strings, is its extensivity in the fivebrane worldvolume, which discourages a pictorial interpretation in terms of a stringy random walk \refs\rcoiled.  
 
 In Section 5 we analyze the thermodynamical balance between the horizon and the radiation in the tube and we find, in particular, the precise regime in which \entkut\ is a good qualitative approximation.
 As a preview of this analysis, notice that  the term \radfe\ resolves the `plateau' degeneracy of \fenn\ in the canonical ensemble. The one-loop corrected free energy at fixed temperature $T=T_H$ becomes a monotonically increasing function of the horizon coordinate, and is  minimized at the lowest possible value of
 $z_0$. This suggests that the canonical ensemble of the system with a {\it finite} tube cutoff is dominated by a pure radiation phase in the tube, i.e. the horizon is pushed down to the edge of the LST regime. 

 Another point is worth stressing. In this paper we will be mainly concerned with the notion of thermodynamical stability, as diagnosed by the sign of the specific heat in homogeneous and static phases. Local dynamical instability inducing time-dependence of the background, such as the  Gregory--Laflamme instability \refs\rglaflamme, is known to be associated with black-brane metrics featuring negative specific heat \refs\rreall. The corresponding connection with the dynamics of LSTs was also made in \refs\rrangamani.  In our analysis, the instabilities are triggered by quantum effects such as a radiation bath, and the analogous dynamical instability associated with these states would be the Jeans instability. These dynamical details, although very interesting, will not be addressed here, since we limit ourselves to the discussion of thermodynamical quantities estimated over translationally invariant (in space and time) states. It is, however, interesting to quote the value of the Jeans length $\ell_J$ in the vicinity of the horizon. On dimensional grounds we have $\ell_J^2 \sim (G_{\rm N} T_H^{10})^{-1} \sim e^{2\phi(r_0)} \ell_s^8 \,T_H^{10} \sim N^\alpha \,R^2 \,E_b /E_H$. This quantitity is always larger than $R^2$, and diverges linearly with the energy of the black brane. Hence the Jeans lifetime of the system increases as we dwell deeper into the Hagedorn regime of the LST.

\newsec{The nonrelativistic string gas on the tube}

\noindent

 The scaling of the expressions \radfe\ and \rades\ suggests a seven-dimensional gas of massless radiation  in the tube. However, the precise coefficient $C_I $, calculated in \refs\rkut, does not correspond to a true massless gas. In fact, it turns out that the radiation gas in the tube has an energy gap $1/R = 2\pi T_H$. This means that the part of the entropy  scaling extensively in $\Delta z$ behaves as that of a nonrelativistic gas at a temperature close to the mass threshold. In this section we explain this nonrelativistic character of the string gas in some detail, as usually this particular feature is not emphasized, but will nevertheless play a key role in our discussion.

 \subsec{The spectrum}

\noindent

The spectrum in the tube can be derived exactly for the whole tower of string excitations, since
this background can be realized as an exact conformal field theory\foot{In this paper we consider mostly the limit in which the geometrical picture of the background is valid, in the $N\gg 1$ limit. Exact conformal field theories may better probe the string-scale physics. In particular, it is possible that there is a transition at the borderline case $N=2$ to a system where the bulk-boundary   partition is better defined (c.f. \refs\rgiveonkrs).} on the worldsheet (c.f. \refs\rchs). The tube geometry
is  ${\bf R}^{1+5} \times {\bf R}_z \times {\bf S}^3$, with ${\bf R}_z$ given by a linear-dilaton CFT in the $z$ direction, with worldsheet field $Z$. The ${\bf R}^{1+5}$ factor is the standard free conformal field theory of six bosons ${Y^a}$ with the corresponding superpartners, and the ${\bf S}^3$ is realized as a $SU(2)_k$ current algebra, with $k=N-2$. The worldsheet fermions of the WZW model are free, so that the complete conformal field theory on the worldsheet is the same as the flat ten-dimensional Minkowski background, except for the replacement of the $c=4$ free conformal field theory of four bosons by the
product of the $SU(2)_k$ WZW model and the linear dilaton on the $Z$ field, with energy-momentum tensor
$T = -\shalf \pt Z \pt Z + \shalf Q \pt^2 Z$, $Q=\sqrt{2/N}$ (in $\alpha'=2$ units). The conformal operators of this sector are oscillator descendants of $\exp(b Z)$, with weight $-\shalf b(b+Q)$. Delta-function normalizable, propagating states in the $z$ direction, have $b = ip_z -Q/2$, so that the weight can be written as $\shalf p_z^2 + Q^2 /8$.  Acting on vertex operators of the form
$$
{\cal O}(\pt Y, \pt Z, J, \psi) \;e^{ip_a Y^a} \,V_{j} \,e^{b Z}\;,
$$
with $V_j$ a WZW primary and ${\cal O}$ a polynomial in derivatives of embedding fields, WZW currents and worldsheet fermions, the Virasoro generator on physical (transverse states) is given by
\eqn\virgen{
L_0 = -\half \omega^2 + \half {\vec p}^{\;2}  + {j(j+1) \over N}  +\half p_z^2 + {Q^2 \over 8}  + {\rm Osc}^\perp\;,}
with ${\rm Osc}^\perp$ denoting the transverse oscillators of all worldsheet  fields and $p^a = (\omega, {\vec p}\,)$ is the energy-momentum along ${\bf R}^{1+5}$. Adding the right-movers we obtain the seven-dimensional mass shell condition
$$
L_0 + {\bar L}_0 - a - {\bar a} = \shalf (- \omega^2 + {\vec p}^{\;2}  + p_z^2 + M^2 ) =0
$$
with $a, {\bar a}$ the normal ordering constants ($a=1/2$ in a quiral Neveu--Schwarz  sector and $a=0$ in a Ramond sector). This condition must be supplemented by the level matching of left and right moving oscillator numbers. Restoring the $\alpha'$ dependence, we obtain the {\it seven-dimensional} mass formula
\eqn\massf{
M^2 = {4j(j+1) \over R^2} + {1\over R^2} + {2\over \alpha'} \left({\rm Osc}^\perp +{\overline {\rm Osc}}^\perp -a-{\bar a} \right),}
 with $R^2 = \alpha' N$. The GSO projection is the same as in the ten-dimensional type IIA string, and   removes the pure zero-mode state in the NS sector. Hence, we see that the tube has a seven-dimensional massive spectrum, with minimum mass equal to $1/R$.  This spectrum is exactly the same as in the ten-dimensional IIA theory in Minkowski space, with the replacement of four flat dimensions contributing
a $p_4^2$ to the dispersion relation, by the ${\bf S}^3$ Casimir and the dilaton shift,
\eqn\subs{
p_4^2 \longrightarrow {4j(j+1) \over \alpha' N} + {1\over \alpha' N}\;.
}

\subsec{The string free energy}

\noindent

This simple fact allows us to reconstruct the free energy  computed in \refs\rkut. The modular-invariant form of the free energy can be written as
\eqn\fremodin{
I^{(1)}_{\rm tube} = \int_{F_0} {d^2 \tau \over \tau_2^2}  \, \CI_{10} (\tau, {\bar \tau}) \; \Upsilon (\tau, {\bar \tau})\;,}
where $F_0$ is the fundamental domain of the torus modular group, $\CI_{10} (\tau, {\bar \tau})$ is the integrand of
the free energy of the ten-dimensional type IIA string theory in Minkowski space, and $\Upsilon$ is the ratio of partition functions
\eqn\ratio{
\Upsilon(\tau,{\bar \tau}) = \left|\CZ_{c=1} (\tau)\right|^{-8} \,\left|\CZ_Z (\tau)\right|^2 \, \CZ_{SU(2)} (\tau,{\bar \tau})\;,}
expressing the substitution \subs. Each chiral partition function in \ratio\ is of the form
$$
Z(\tau)_{\rm CFT} = \Tr_{\CH_{\rm CFT}} \;\;q^{L_0 - c/24}
$$
with $q= e^{2\pi i \tau}$. In particular, $\CZ_{c=1}$ is the partition function of a free scalar, $\CZ_Z$ is
the partition function of the Liouville field $Z$ and the remaining factor is the diagonal modular invariant partition function of the WZW model.  In fact, only three factors of a $c=1$ system are necessary in \ratio, since the partition function of the Liouville field $Z$ with background charge $Q=\sqrt{2/N}$ is exactly given by one $c=1$ factor (c.f. \refs\rberkleb), the dilaton shift $Q^2 /8$ in the conformal weights being partially offset by the usual additive normalization of the worldsheet Hamiltonian $-c_Q/24 = -1/24 - Q^2 /8$. This well-known property of Liouville fields can also be seen from the path-integral representation of the partition function, since the background charge $Q$ enters through the worldsheet coupling to the two-dimensional curvature, $\int d^2 \sigma Q Z R^{(2)}$, which vanishes in the conformal gauge for the genus one worldsheet.

 The fact that the Liouville partition function is only sensitive to $c_{\rm eff} = 1$ seems to imply  that the dilaton shift in the mass spectrum will not be visible in the one-loop free energy of the string gas. However, closer inspection reveals that  the mass gap is to be found in  the WZW partition function. With $SU(2)$ highest weights $L_0 |j\ket = h_j |j\ket$, we have
\eqn\wzws{
h_j - {c_{SU(2)} \over 24} = {j(j+1) \over N} - {3(N-2) \over 24N} = {j(j+1) \over N} - {3 \over 24} + {1\over 4N}\;.}
In this expression, the factor of $3/24 = 1/8$ is the standard Virasoro shift for 3 bosons, whereas the excess $1/4N = Q^2 /8$ is the term responsible for the mass gap. All in all, the product of the Liouville CFT with $Q=\sqrt{2/N}$ and the WZW model at level $N-2$ is a $c=4$ conformal field theory with
a gap $1/4N$ in the weights of primary fields.

Standard manipulations of the modular invariant partition function
 (reviewed in \refs\rkut) provide a proper-time representation of the free energy as a sum over the field excitations that appear in the spectrum,
 \eqn\propert{
 I^{(1)}_{\rm tube} = -\beta\,V_5 \, \Delta z \,\int_0^\infty {d\tau_2 \over 2\tau_2}  (4\pi^2 \alpha' \tau_2)^{-{7\over 2}} \sum_{f,j} \sum_{\ell \in {\bf Z}} (-1)^{(\ell+1) F_f} \,\exp\left(-\pi \alpha' \tau_2 M_{f,j}^2-{\ell^2 \beta^2 \over 4\pi \alpha' \tau_2}\right)\,.
 }
 In this expression, $(-1)^{F_f}$ is the spacetime fermion number of the field excitation $f$,  with mass squared $M_{f,j}^2$ given by formula \massf, namely the sum over discrete energy levels on the ${\bf S}^3$ is explicitly denoted in \propert, whereas the momenta along the world-volume directions as well as the `tube' direction have been already integrated out. 
 The proper-time integrals can be expressed in terms of modified Bessel functions:
 \eqn\modb{
 I^{(1)}_{\rm tube} = -2 \beta\,V_5 \,\Delta z \,\sum_{f,j} \sum_{\ell=1}^\infty (-1)^{(\ell+1) F_f} \left({M_{f,j} \over 2\pi \beta \ell}\right)^{7\over 2} \; K_{7\over 2} (\ell \beta M_{f,j})\;.}
 Alternatively, we can sum  over winding numbers $\ell$ in \propert\  and arrive at the familiar form of
  the free energy in the ideal gas approximation
\eqn\idgas{
I^{(1)}_{\rm tube} = V_5 \,\Delta z \,\sum_{f,j} (-1)^{F_f} \int {d{\vec p}\, d p_z \over (2\pi)^6} \log\left(1- (-1)^{F_f} e^{-\beta \omega_{f,j}}\right)\;,
}
where $\omega_{f,j} = \sqrt{{\vec p}^{\;2} + p_z^2 + M_{f,j}^2}$.
As expected, we find an ideal gas with full extensivity in the $z$ coordinate. Since $M_f \geq 1/R$, the thermal gas is effectively non-relativistic for temperatures $TR<1$. In particular, this is the case at the nominal LST temperature $T_H = (2\pi R)^{-1}$. At low temperatures, we can then approximate the free energy by its standard nonrelativistic form, obtained from \modb\ by keeping only the $\ell= 1$ term:
\eqn\nonrelf{
I^{(1)}_{\rm tube} \approx  -V_5 \,\Delta z\,N_0 \,\left({M_0 \over 2\pi \beta}\right)^{3} \;\exp(-\beta M_0)\;,}
with $M_0 = 1/R$ and  $N_0 =256$ in our case, the total number of field excitations at the lower mass level.

 \newsec{Bulk radiation versus LST holography}

 \noindent

 The extensivity of the bulk radiation in `the tube' looks threatening for a standard holographic interpretation of the one string-loop corrections.
 In order to emphasize this  issue we pause to discuss some general aspects of the relation between bulk thermal extensivity and holography (c.f. \refs\rus).

 We begin by recalling the well known situation in ${\rm AdS}_{d+1}$ spaces, with a metric of the form
 $$
 ds^2 = {r^2 \over \CR^2} (-dt^2 + dx_{d-1}^2 ) + \CR^2 {dr^2 \over r^2}
 $$
 in a Poincar\'e coordinate patch.   A black $(d-1)$-brane horizon at temperature $T $ cuts off
 the AdS space at $r_0 = 2\pi \CR^2 T$. The entropy stored `classically' on this horizon scales holographically, i.e. as in a $d$-dimensional CFT with a number of particle degrees of freedom, $N_{\rm eff
}$, proportional to the inverse  Newton's constant in the bulk:
 $$
S_{\rm BH} = {V_{d-1} \,r_0^{d-1} \over 4 G_{d+1} \, \CR^{d-1} } ={(2\pi \CR)^{d-1} \over G_{d+1}} \,V_{d-1}
\,T^{\,d-1} \equiv N_{\rm eff} \;V_{d-1} \; T^{\,d-1}\;.
$$
At the same time, the entropy  stored in radiation, far above the horizon and  up to infinity, can be estimated to be
\eqn\entwkb{
S_{r} \sim \int_{r>r_0}  d{\vec V}  \left({T \over \sqrt{-g_{tt}}}\right)^{d}
\;,}
where $d{\vec V}$ is a spatial proper-volume element and $T /\sqrt{-g_{tt}}$ is the local red-shifted temperature. A more geometrical characterization of this quantity can be obtained in the Euclidean formalism,  by recalling that
thermal radiation in static, curved spacetimes, is extensive in the so-called {\it optical manifold} (c.f. \refs{\roptical, \rgiboptical}),  a conformally rescaled spacetime with metric $g_{\mu\nu} / |g_{tt}|$, after continuation to the Euclidean signature and compactification of time, $it \equiv it + \beta$, on a circle of length $\beta = 1/T$. Performing the calculation, we find
$$
S_{r} \sim  C_S \,T^{d+1} \;{\rm Vol} \,({\rm AdS}_{\beta} )_{\rm optical} \sim V_{d-1}\; T^{\,d-1}
\;,$$
 a result that shows
no vestiges of `radial extensivity'. The scaling ensures its holographic interpretation as the first $1/N_{\rm eff}$ correction to the entropy of the CFT in $d$ dimensions. In fact, this holographic scaling generalizes to all metrics of D$p$-branes with
standard  holographic behaviour, i.e. $p<5$, with the substitution $d=p+1$. The first failure occurs precisely for $p=5$, the metric S-dual to \hololst.

\subsec{Effective potentials}

\noindent

An alternative way of exhibiting the delicate balance between holography and bulk radiation extensivity is to study  the spectrum of normal frequency modes associated to the time coordinate of some global static frame, $\omega = -i \pt_t$.
 It will be useful to   extend the  discussion to metrics of the general form
\eqn\genm{
ds^2 = -f(r) \,dt^2 + {dr^2 \over g(r)} + \sum_i \varrho_i (r)^2 \,ds_i^2\;,
}
where $ds_i^2$ are metrics of $d_i$--dimensional Riemannian spaces, with warping factors $\varrho_i^2$. Let us consider a minimally coupled scalar field from the NS-NS sector, with action
\eqn\gens{
S_\Psi = -\shalf \int d^D x \,\sqrt{-g} \;e^{-2\phi} \,\left( g^{\mu\nu} \pt_\mu \Psi \pt_\nu \Psi + m^2 \Psi^2 \right)\;,}
where the metric and the dilaton are treated as classical backgrounds, and furthermore we assume that the dilaton profile is only a function $\phi(r)$ of the radial coordinate. The nontrivial dilaton profile can be eliminated from the kinetic term by the redefinition $\Psi = e^\phi {\widetilde \Psi}$, resulting in an action of the form \gens\ with an  effective mass shift
\eqn\rulephi{
m^2 \rightarrow {\widetilde m}^{\,2}  = m^2  + (\nabla \phi)^2 + \nabla^2 \phi
\;,}
after a total derivative is discarded from the action. In order to reach canonical normalization for the normal frequency modes in the bulk, it is convenient to define a generalized `Regge--Wheeler'  radial coordinate,
\eqn\red{
dz = {dr\over \sqrt{f(r)g(r)}}\;,}
 casting the original metric in the form
\eqn\genmrw{
ds^2 = f(z)\,\left(-dt^2 + dz^2 \,\right) + \sum_i \varrho_i (z)^2 \,ds_i^2\;,}
where all functional dependencies on $z$ are defined through the change of variables \red. Then, the further field redefinition
\eqn\redef{
{\overline \Psi} = \left( \prod_i \varrho_i (r)^{d_i} \right)^{1/2}  \;{\widetilde \Psi}\;,}
yields, after partial integration
\eqn\redac{
S_\Psi = -\shalf \int dt dz \prod_i dV_i \; {\overline \Psi} \,\left( \pt_t^2 - \pt_z^2 + V_{\rm eff} (z) \right)\,{\overline \Psi} \;,
}
where $dV_i$ stands for the volume element if the $i$-th manifold with metric $ds_i^2$. The effective potential reads
\eqn\effpot{
V_{\rm eff} (z) = \shalf \,\pt_z^2 \,\log(\varrho_\Pi ) + \sfourth \left[\pt_z \,\log (\varrho_\Pi )\right]^2 + f(z)
\left( {\widetilde m}^{\,2} + \sum_i {\Delta_i \over \varrho_i (z)^2} \right)\;,}
where $\varrho_\Pi = \prod_i \varrho_i^{d_i} $ is the product of all warping factors  and $\Delta_i$ is the
Laplace operator on the $i$-th warped submanifold.

 Entirely similar effective potentials can be constructed for general fields of arbitrary mass and spin, after  due  attention is paid to some subtleties. For example, in the case of the dilaton-graviton system in the NS-NS sector, the field-redefinition $\Psi \rightarrow e^\phi \,{\widetilde \Psi}$ that removes the dilaton prefactor from the action affects only the classical part of the dilaton field. The fluctuating dilaton field mixes with the trace of the metric and this effect must be disentangled in order to find the spectrum of physical eigenfrequencies (for example, by transforming to the Einstein frame). In all cases, the effect of the classical dilaton background  is to induce a  shift  of the effective mass  according to the rule
 \rulephi. In particular, for a linear dilaton background, $\phi(z)  =  -z/R$, we have a genuine shift of the mass-squared by $1/R^2$.   In the case of Ramond-Ramond $p$-forms and spacetime fermions, the action  is simpler when written without the dilaton prefactor in the Lagrangian. In these cases, the mass-shift is seen in the supergravity approximation as induced by nontrivial fluxes, such as the NS flux on the ${\bf S}^3$ in the case of NS5-branes.

For each field excitation, the  normal frequency spectrum is identified with the spectrum of the Schr\"odinger operator
\eqn\ops{
\omega^2 = -\pt_z^2 + V_{\rm eff} (z)\;,}
and the one-loop free energy follows from $\omega$ as
\eqn\onelo{
I^{(1)} = \Tr_{\CH_{\rm ph}} \,(-1)^F \;\log\left(1-(-1)^F \,e^{-\beta \omega} \right)\;,
}
where the trace over the Hilbert space of physical states includes both spacetime and internal quantum numbers, as well as field types. The approximation  \entwkb, based on the assumption of local extensivity in terms of the  red-shifted temperature, can be substantiated by a  WKB approximation of \onelo\ (c.f. \refs\rwkbs\ and Appendix A).  To be more specific, we can work backwards from \onelo\ as in the previous section to obtain the supergravity analog of \propert,
\eqn\propgen{
I^{(1)} = -{\beta \over2}  \int {d\tau_2 \over \tau_2} \sum_f \sum_{\ell \in {\bf Z}} (-1)^{\ell F_f}\; \Lambda_f (\tau_2)\; \exp\left(-{\ell^2\beta^2
\over 4\pi\alpha' \tau_2}\right)\;,
}
where
\eqn\vacint{
\Lambda_f (\tau_2) = {(-1)^{F_f} \over \sqrt{4\pi^2 \alpha' \tau_2}} \;\Tr_{\CH_f} \,\exp\left(-\pi\alpha' \tau_2 \,\omega_f^2\right)\;.}
We can approximate the trace over the quantum numbers of the field excitation $f$ by an integral, with the WKB approximation to the Schr\"odinger spectrum \ops. The result is (see Appendix A)
\eqn\resl{
\Lambda_f (\tau_2) \approx  {(-1)^{F_f} \over 4\pi^2 \alpha'  \tau_2} \sum_{\Delta_i} \int dz \;e^{-\pi\alpha' \tau_2 \,V_{\rm eff} (z)}\;,}
showing  that the effective potential $V_{\rm eff}$ substitutes  $M^2_f$ in \propert. \foot{The different power of $4\pi^2 \alpha' \tau_2$ in \propgen\ and  \resl\ with respect to \propert\ can be traced back to the fact that, in the notation adopted in this section, the collective label for the harmonic quantum numbers, $\Delta_i$, includes the world-volume momenta as well.}  In this form, we can track the contribution of different regions of the warped geometry \genmrw\ to the radiation free energy.  In particular, if $V_{\rm eff}$ shows a plateau, there is strict extensivity in the Regge--Wheeler coordinate and the free energy is proportional to $\Delta z$.

The characteristic holographic behaviour (AdS spaces) is quite the opposite, since
the Regge--Wheeler coordinate $z$ is bounded as $r\rightarrow \infty$. From the definition \red\ and \genmrw\ 
we find $f(r) = r^2 /\CR^2$ and $z=z_\infty - \CR^2 /r$ for the particular case of an AdS metric. Here, $z_\infty$ is a finite integration constant that arises when integrating \red, setting the bound on the $z$ coordinate.  The potential is then proportional to $f(z) \sim \CR^2 /(z-z_\infty)^2$, showing a finite range and an infinite wall at $z=z_\infty$. This is a very graphical demonstration of the `box-like' nature of AdS spaces, despite their non-compactness in the radial direction. For the metrics associated to near-horizon D$p$-branes with $p<5$  we find instead
\eqn\ndp{
{r \over R_{{\rm D}p}} = \left({2 \over 5-p} \right)^{2 \over 5-p} \;\left({R_{{\rm D}p} \over z_\infty - z} \right)^{2 \over 5-p}\;,}
where $R_{{\rm D}p}$ denotes the charge radius of the D$p$-brane. This expression shows that  all holographic throats of  D$p$-branes with $p<5$ have finite range in the $z$ coordinate. In the $p=5$ case, we find the first marginal failure of holography, as $z$ becomes logarithmically related to $r$ and thus   unbounded  as $r\rightarrow \infty$. Strictly speaking, the  asymptotic behaviour of the potential for $p>3$  must be studied in the S-dual system, since the local string coupling grows at large $r$ in all those cases. In particular, for D5-branes we are forced to study the corresponding background of NS5-branes. 

\subsec{Fivebrane potentials}

\noindent

For our main subject in this paper, the NS5-branes, we deal with metrics of the form \hololst\ in the near-horizon regime. In this case, $z$ is precisely the Regge--Wheeler coordinate of this system, and the effective potential can be obtained from the general formula \effpot. In the parametrization \genmrw\ of the metric, we have $f(z)=1$ and two `warped' submanifolds: ${\bf R}^5$ with trivial warping factor, and 
${\bf S}^3$ with constant warping factor $\rho(z)=R$.  Since the eigenvalues of the Laplacian on ${\bf R}^5$ are given by standard momenta ${\vec p}$, we find the potential 
\eqn\efflst{
V_{\rm eff} (z) = {\vec p}^{\,2} + {\widetilde m}^2  + {\Delta_{{\bf S}^3} \over R^2} \;,}
where  ${\widetilde m}^2 = m^2 + 1/R^2$. 
The Schr\"odinger operator \ops\  can be diagonalized by plane-waves in the $z$ direction with momentum $p_z$, and we end up with a frequency spectrum
\eqn\dispr{
\omega^2 = {\vec p}^{\;2} + p_z^2 + m^2 +{1\over R^2} + {\Delta_{{\bf S}^3} \over R^2} \;,}
Showing that a minimum  mass-gap $1/R$ persists, even for the massless ten-dimensional multiplet, with $m=0$.

If we include the asymptotically-flat region of \fullten, the mass shift $m^2 \rightarrow m^2 + 1/R^2$ on the tube eventually turns off for $r\gg R$, or $z\gg z_R \sim R\log(1/g_s)$. At small $r$,
the potential \efflst\ is formally valid across the strong-coupling transition at $r\sim r_s = g_s R$ (i.e. $z\sim z_s=0$), provided the field $\Psi$ does not represent an excitation with momentum in the eleventh dimension. At the `LST threshold', $z=z_{H}\sim -R\log N$, corresponding to $r\sim r_H = g_s R /N$ in the type IIB theory,  the supergravity approximation breaks down and we must match the system to a six-dimensional Yang--Mills theory. Therefore, $z=z_{H}$ is an explicit boundary for the type IIB effective potential. In the type IIA case, we can match the system to the ${\rm AdS}_7 \times {\bf S}^4$ background within the accuracy of the supergravity approximation. Hence, one glues the AdS effective potential to the LST effective potential at $z=z_{H}= -R\log \sqrt{N}$, resulting in a complete potential shown in Figure 1.

\fig{\sl   Effective potential of the complete  IIA fivebrane background. The  eleven-dimensional low-energy AdS part is glued at $z\sim z_H$ onto the  CHS background, composed of the `tube' plateau opening to the ten-dimensional asymptotically  flat region far away from the fivebranes. The dashed lines represent the pure ${\rm AdS}_7$ potential  blowing up at $z_\infty$, which can be determined by matching the AdS potential to the plateau of height $m^2 +1/R^2$ when $z \sim z_H$. This fixes
$z_\infty - z_H \sim \shalf R \sqrt{35}$ in the $m\rightarrow 0$ limit.}{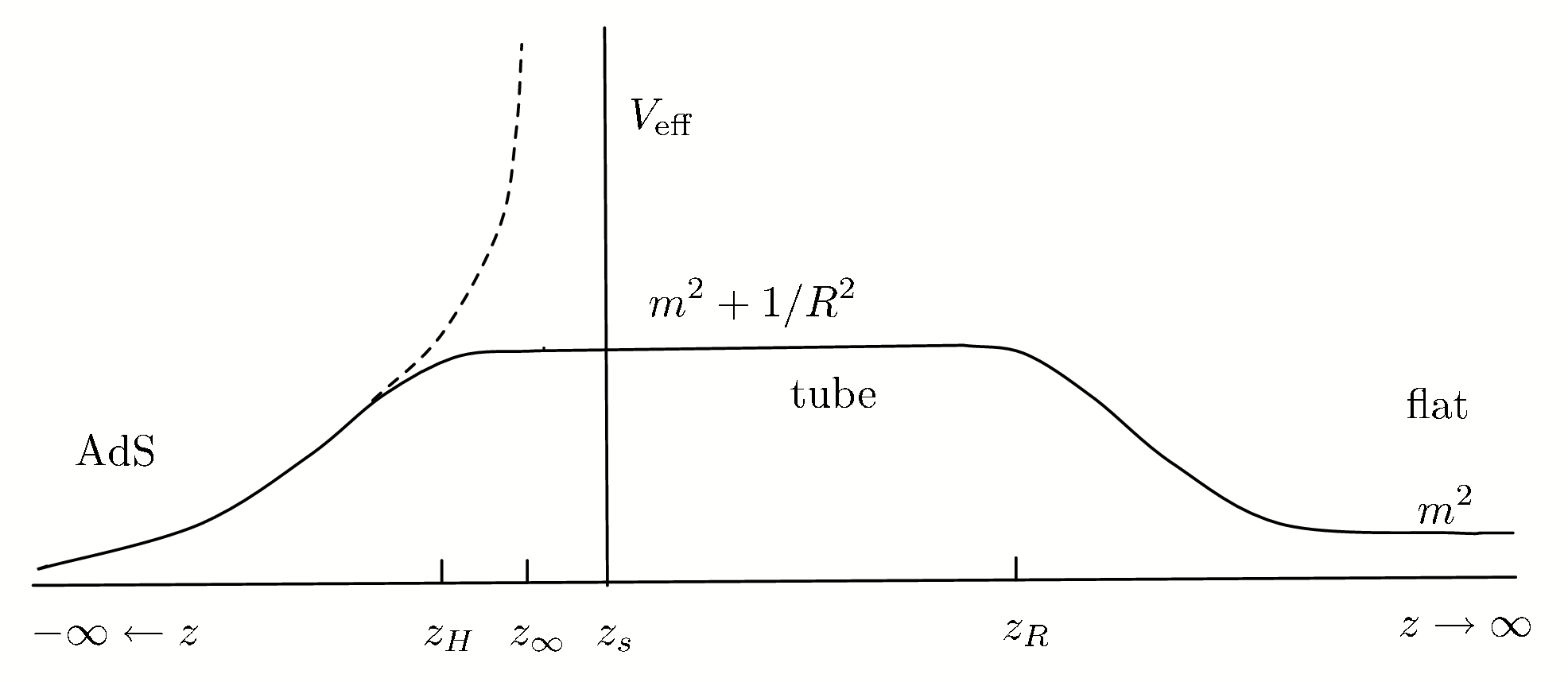}{5truein}

The effective potential in Fig. 1 gives an interesting angle on the workings of   the  LST
holography.  States localized at $z\ll z_{H}$ represent the excitations of the low-energy $(2,0)$ CFT. These excitations are separated from the  ten-dimensional Minkowski region by a barrier of height
 $1/R^2$. In the LST decoupling limit we have  $g_s \rightarrow 0$, which is equivalent to making the barrier infinitely long. Insistence on keeping the LST scale $T_H$ fixed will result in  the barrier  retaining a  {\it finite height} in this limit (unlike the analogous situation for D$p$-branes with $p<5$) and thus excitations with energy above the $T_H$ threshold will propagate freely in the $z$ direction. For this reason, it is usually stated that the  `tube' region at $z>z_{ H}$ cannot be removed from the holographic dual and the continuum spectrum of momentum states with energy above $T_H$ is to be considered as part of the LST physical spectrum (c.f. \refs{\rmaldastro, \rminwallasei}). A discussion of these tube degrees of freedom in the matrix model definition of LST appears in \refs\rdlcqt. 

 \newsec{Thermodynamics of a cutoff LST model}

 \noindent

  We have argued in  Section 2 that the thermodynamics of the LST, as defined by its gravity dual,
  is afflicted by an infrared divergence coming from the entropy of radiation `in the tube' (the low-energy part of the string one-loop corrections).  On physical grounds, this contribution cannot be neglected, since the black-brane horizon injects radiation in the tube at a finite rate  (see \refs{\rmaldastro, \rminwallasei}). In fact, the massless radiation in the tube dominates over the short-distance loop corrections provided the tube is sufficiently  long.
   In this section we analyse this
 problem by simply imposing a cutoff in the length of the tube.    In this way, the entropy and free energy of the radiation, \rades\ and \radfe\ contribute a finite amount, and we can study whether it comes to equilibrium with the black-brane horizon.

We can idealize the system as a `piston', modeling the horizon in thermal contact with radiation in `the tube', reaching out to a total length $\Delta z = z_\Lambda - z_0 = \log(r_\Lambda /r_0)$, where $z_\Lambda$ is the coordinate of the tube boundary (c.f. Fig 3). It will be useful to define a  dimensionless coordinate
$x$, proportional to $z$ and such that $x=0$ at the LST threshold $z_H$. The relation between the different radial coordinates is then
\eqn\relcor{x\equiv {z - z_H \over R} = \log(r/r_H)\;,}
where $r_H = g_s R \,N^{-2+\alpha/2}$ and $z_H = R \log(N^{-2+\alpha/2})$. Recall that $\alpha =2$ in the type IIB model and $\alpha=3$ in the type IIA model.
 
  The tube cutoff sits at $x=x_\Lambda$, the maximal length of the tube being $z_\Lambda - z_H =R\,x_\Lambda$, and the fraction of tube length available for radiation being $z_\Lambda - z_0 =     R\,(x_\Lambda -x_0)$ (see
Fig. 3).
The boundary of the tube at $x=0$ is assigned the internal degrees of freedom of the low-energy systems, i.e. the six-dimensional $SU(N)$ Yang--Mills theory in the IIB model, and the rank $N$, $(2,0)$
CFT in the IIA model. In this last case, we can actually continue the geometrical description to negative values of the $x$ coordinate, by matching the tube to the ${\rm AdS}_7 \times {\bf S}^4$ geometry
of the infrared fixed point.  In either case, we can model the system as a tube with internal degrees of freedom at the lower, $x=0$ boundary, worth an entropy and internal energy 
\eqn\lowbg{
S_{\rm IR} \sim N^\alpha \,V_5 \,T^5 \;, \qquad E_{\rm IR} \sim N^\alpha \,V_5 \,T^6\;,}
where $T$ is the temperature of the infrared boundary. At the LST threshold, $T=T_H$ and we denote
these quantitites by $E_H = N^\alpha V_5 T_H^6$ and $S_H = \beta_H E_H$.  The resulting cutoff LST model can be parametrized by the threshold internal energy, $E_H$, the intrinsic length scale $\beta_H \sim R$,  and the dimensionless tube cutoff, $x_\Lambda$.

We approximate the system by two components in thermal contact: black brane and radiation with free energy
$
\beta F = I= I_b + I_r
$,
 with a fixed temperature $T= 1/\beta$. The black free energy is 
$
I_b = \beta \,E_b - S_b
$,
with $E_b$ denoting the energy of the classical black-brane solution, normalized with respect to the extremal solution, and $S_b$ is the  Bekenstein--Hawking entropy:
\eqn\feen{
E_b (x_0) = E_H \,e^{2x_0}\;, \qquad S_b (x_0) = S_H \,e^{2x_0} = \beta_H \,E_b (x_0)\;.
}
Above the horizon we have a maximally entropic state of radiation
contained in the excluded volume, i.e. in the region $x_0 < x < x_\Lambda$.

\fig{\sl The `piston model' of the LST thermodynamics. The tube extends from the `infrared boundary' at $x=0$,  out to $x=x_\Lambda$. The horizon reaches out to $x=x_0$, and the remaining volume is filled with a maximally entropic state of radiation.}{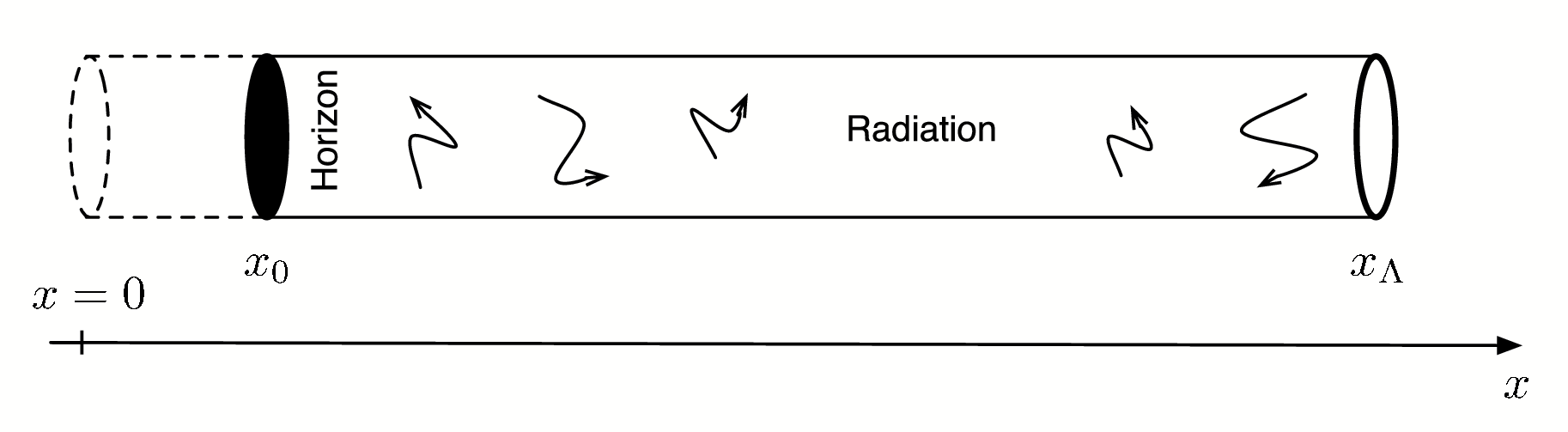}{4truein}

The radiation component has two regimes depending on whether its
temperature is above or below the mass gap of the cylinder $1/R \sim T_H$.
For $T > T_H$ we have a ten-dimensional  relativistic gas with free energy
\eqn\radrel{
\beta F_r  =I_r \sim - V_r \,T^9
\;,}
and effective volume $V_r \sim (x_\Lambda - x_0) \,R^4 \,V_5$,
whereas for $T <T_H$ we have a seven-dimensional  non-relativistic gas of
particles of mass $M \sim 1/R$, as in \nonrelf,
\eqn\readnrel{
I_r \sim - V_r
\,T^{6} \; (TR)^{-3} \; e^{-\beta/R}
\;,}
where now  the radiation volume is given by $V_r \sim (x_\Lambda - x_0) \,R\,V_5$.  Notice that both
\radrel\ and \readnrel\ match within $O(1)$ constants at $T\sim T_H$. In what follows we neglect such
$O(1)$ coefficients.

Our purpose is to investigate the equilibrium conditions between horizon and radiation degrees of freedom, depending on external parameters: the temperature in the canonical ensemble, and the total energy in the microcanonical ensemble. In this analysis, we bring the horizon `off shell', as we decouple the value of the temperature from the value of the horizon radius. In the Euclidean formalism, we would be discussing metrics of off-shell black-branes, with a conical defect at $r=r_0$.

\subsec{Canonical balance}

In the canonical ensemble at temperature $T\leq T_H$, the total free energy of the system is
\eqn\radp{
I(x_0) = I_b (x_0) + I_r (x_0) \sim
 (\beta-\beta_H) \,E_H \,e^{2x_0} + (x_0 - x_\Lambda) \, V_5 \,T^3\, T_H^2 \,
e^{-\beta/R}
\;,}
which is minimized at $x_0=0$. Hence, at temperatures $T\leq T_H$, the horizon surrenders its energy to the Hawking radiation in the tube, and recedes to the boundary with infrared degrees of freedom.  The subsequent balance takes place between the degrees of freedom at the IR boundary,
with free energy
$$
I_{\rm IR} \sim - N^\alpha \,V_5 \,T^5\;,
$$
 and the non-relativistic radiation in the tube.  For a given temperature $T<T_H$,
the radiation in the tube dominates provided the cutoff satisfies
$$
x_\Lambda \gg N^\alpha \,(T/T_H)^2 \,e^{\beta/R}\;.
$$
Equivalently, for a given tube cutoff, the system shows six-dimensional CFT behaviour for temperatures
$T< T_{\rm t}$, and seven-dimensional behaviour for $T_{\rm t} < T<T_H$, where the critical temperature for `tube domination' is
\eqn\tubedom{
T_{\rm t} \sim {T_H \over \log (x_\Lambda /N^\alpha)}\;.
}

The canonical ensemble at $T>T_H$ is more interesting. The free energy
is now
\eqn\htfe{
I(x_0) \sim (\beta-\beta_H)\,E_H \,e^{2x_0} + (x_0 - x_\Lambda) \,V_5 \,R^4 \,T^9\;,
}
with $\beta-\beta_H <0$.
For any $T>T_H$, the removal of
the infrared cutoff  $x_\Lambda
\rightarrow \infty$ sees the global minimum of $I(x_0)$  occurring at $x_0=x_\Lambda$, i.e. with the largest  possible black brane taking up all the available
free energy. For $0< T-T_H \ll T_H$ and very large $x_\Lambda$, both $x=0$
and $x=x_\Lambda$  are `local' minima, whereas
\eqn\maxf{
x_m = \shalf\, \log \left({R^4 V_5 T^9 \over (\beta_H -\beta) E_H} \right) \sim
 \shalf \,\log \left({(T/T_H)^9 \over
N^\alpha \,(1-T_H /T)} \right)\;.
}
is a local maximum characterized by the unstable
balance  of ``pressures" $dI/dx$
between the two phases. Taking $x_\Lambda \rightarrow \infty$ at fixed $T>T_H$
gives an all-black global minimum, whereas the opposite is true in the
$T\rightarrow T_H$ limit at fixed $x_\Lambda$. More precisely, given $x_\Lambda$, the global minimum of $I(x_0)$ in the interval $0<x_0 < x_\Lambda$ is at $x=x_\Lambda$ (black dominance) provided
the temperature rises above $T_*$, with
\eqn\limt{
T_* = T_H \left(1+N^{-\alpha} \,x_\Lambda \,e^{-2x_\Lambda}\right)\;,
}
a rather small increment for large $x_\Lambda$.
In the high-temperature limit the local maximum migrates to large values of $x$, which means that
there is a large region where the black brane sheds energy to the radiation
component, even if the absolute minimum of the free energy is the all-black
configuration.

\fig{\sl Canonical free energy as a function of the horizon position in the interval  $0\leq x_0 \leq x_\Lambda $, for different values of the temperature, ranging form sub-Hagedorn, $T<T_H$, to  temperatures well above $T_H$. The full-radiation state, $x_0 =0$, is the global minimum for $T\leq T_H$. A local full-black minimum at $x_0 = x_\Lambda$
develops soon after the temperature rises above $T_H$. This minimum equilibrates with the full-radiation minimum at temperatures $T\sim T_*$, and quickly becomes dominant for $T>T_*$. This is a typical Landau picture of a first-order phase transition.}{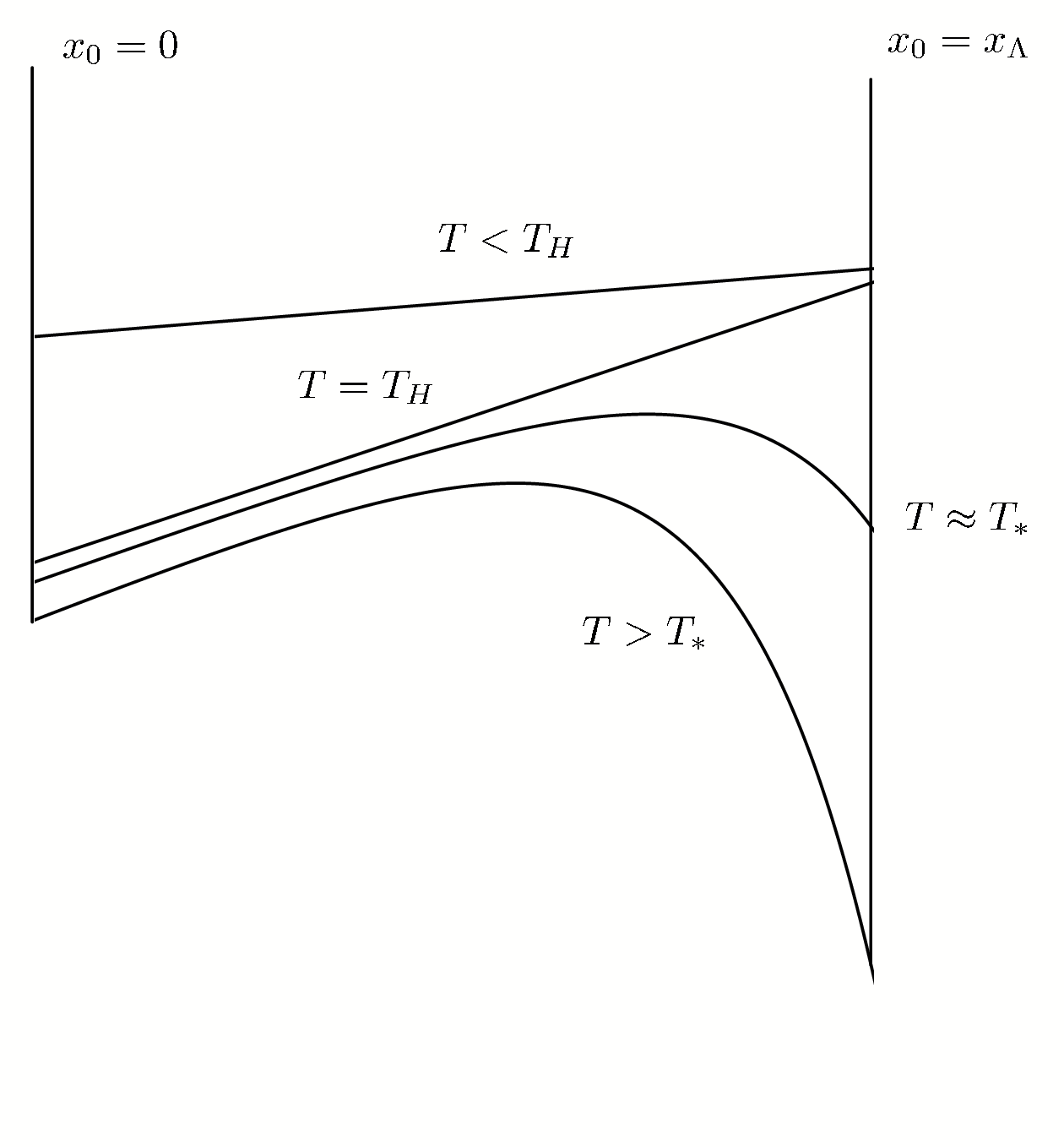}{3.5truein}

We conclude from this analysis that the string loop corrections, in the form of a radiation bath, induce a first-order phase transition similar to the Hawking--Page transition in AdS spaces.
For $T<T_{\rm t}$ we have the standard IR  six-dimensional
 CFT thermodynamics, but the presence of the
long tube with a finite gap opens up a transient $T_{\rm t} < T < T_H$ in which the free energy
is dominated by a seven-dimensional non-relativistic gas. At temperatures only slightly above $T_H$,
the system makes a discontinuous jump to a phase with a large black brane saturating all the available
space of the tube. In the limit where we remove the tube cutoff at fixed $T$ and $T_H$, the system
degenerates: it is dominated by a nonrelativistic seven-dimensional gas for all $T<T_H$, whereas it is unstable against the unlimited growth of a horizon for $T>T_H$.

This result suggests that there is no truly Hagedorn regime, except for energies close to the cutoff scale.
In turn, this implies that the fate of the Hagedorn phase in LST systems is linked to the particular physical implementation of the cutoff.

\subsec{The microcanonical way}

\noindent

The canonical analysis confirms that the thermodynamical behaviour of the system is rather subtle, showing either runaway behaviour or seven-dimensional extensivity upon removal of the radial cutoff. It is then interesting to confirm these findings in  the more fundamental microcanonical ensemble.

Let us consider the typical state of the system at a given total  energy $E$. We adopt again the approximation of three non-interacting components: the low-energy degrees of freedom at the IR boundary $x=0$, with a six-dimensional CFT scaling of entropy:
\eqn\ird{
S_{\rm IR}(E_{\rm IR}) \sim N^{\alpha/6} \, V_5^{1/6} \, (E_{\rm IR})^{5/6}\;,
}
the black brane with energy $E_b=E_H \,\exp(2x_0)$ and Hagedorn entropy
$S_b = \beta_H E_b$, and the thermal gas of the gravity multiplet in the tube of length $R\,x_\Lambda$. In the non-relativistic regime,
\eqn\nrel{
S_r (E_r) \sim {\beta_H\,E_r } \;
 \left(\log \left(V_r T_H^7  / E_r \right)+1\right)
\;\;\qquad {\rm for} \;\;\;\;\; E_r \ll   V_r  \,T_H^7
\;,}
with $V_r = (x_\Lambda -x_0) R V_5$. In the relativistic regime,
\eqn\srad{
S_r (E_r) \sim V_r^{1/10} \; E_r^{9/10} \;\;\qquad  {\rm for}\;\;\;\;\;E_r \gg   V_r  T_H^{10}
\;,}
 with effective volume
$V_r = (x_\Lambda - x_0) R^4 V_5$.

 For $E<E_H$ there cannot be a black brane in the tube,  and the balance takes place between the IR degrees of freedom at the IR boundary and the radiation in the tube, which can only be nonrelativistic for these low energies. One finds a threshold energy
\eqn\etbe{
E_{\rm t} = {E_H \over (\log (N^{-\alpha} x_\Lambda))^6}\;
}
 above which the non-relativistic radiation in the tube dominates the entropy. As $E$ reaches the Hagedorn threshold $E_H$, the black-brane component takes over the IR degrees of freedom as the
 lower boundary of the tube sits at the horizon $x_0 >0$. The total entropy in black-brane plus nonrelativistic radiation as a function of the horizon location reads
\eqn\nrelbal{
 S(x_0)/S_H \sim  e^{2x_0} + \left({\cal E} -e^{2x_0} \right) \;\log \,{x_\Lambda
- x_0 \over N^\alpha ({\cal E} - e^{2x_0} )}
\;,}
where we use the notation ${\cal E} \equiv E/E_H$. This function is monotonic and maximized at
$x_0 =0$, which means that, as long as the radiation in the tube remains nonrelativistic, it will always
win over the black brane.

 The band of Hawking radiation domination ends when the temperature of
the radiation is of order $T_H$, at energies or order
\eqn\bthr{
E_* = E_H \;N^{-\alpha}\, x_\Lambda\;.}
For $E>E_*$ we must model the radiation gas as relativistic. The total entropy as a function of $x_0$ is now
\eqn\relbal{
 S(x_0)/S_H \sim e^{2x_0} + N^{-{\alpha\over 10}} \left(x_\Lambda -x_0 \right)^{1\over 10} \;
\left({\cal E} - e^{2x_0} \right)^{9\over10}
\;.}
In the regime of interest, we have  ${\cal E} >N^{-\alpha} x_\Lambda$, and this function  is
 maximized at some equilibrium value $x={\bar x}$ that quickly approaches the maximum possible
 ${\bar x} \rightarrow \shalf \log {\cal E}$.  We can see this in the following figures, showing the
 evolution of the entropy maximum for different values of the ratio $N^\alpha {\cal E} /x_\Lambda$, in the particular case of the type IIA model ($\alpha =3$).

%\vskip 1cm

\fig{\sl The entropy balance between radiation and a black brane reaching out to $x=x_0$ depends on
the dimensionless ratio $\eta = N^{\alpha}{\cal E}/ x_\Lambda$. In this picture we show the entropy curves $S(x_0)$ for  $\eta$ ranging from $10^{4}$ in $(a)$ to $10^{-2}$ in $(d)$, each step decreasing it by a further  factor of $10^{-2}$. Curves $(a)$ and $ (b)$  correspond to the relativistic gas (eq. \relbal) and curve $(d)$ is the non-relativistic gas \nrelbal. Curve $(c)$ corresponds to the crossover regime $\eta \sim 1$ and features a local maximum for some intermediate value in the interval $0<x_0 < \shalf \log{\cal E}$. }{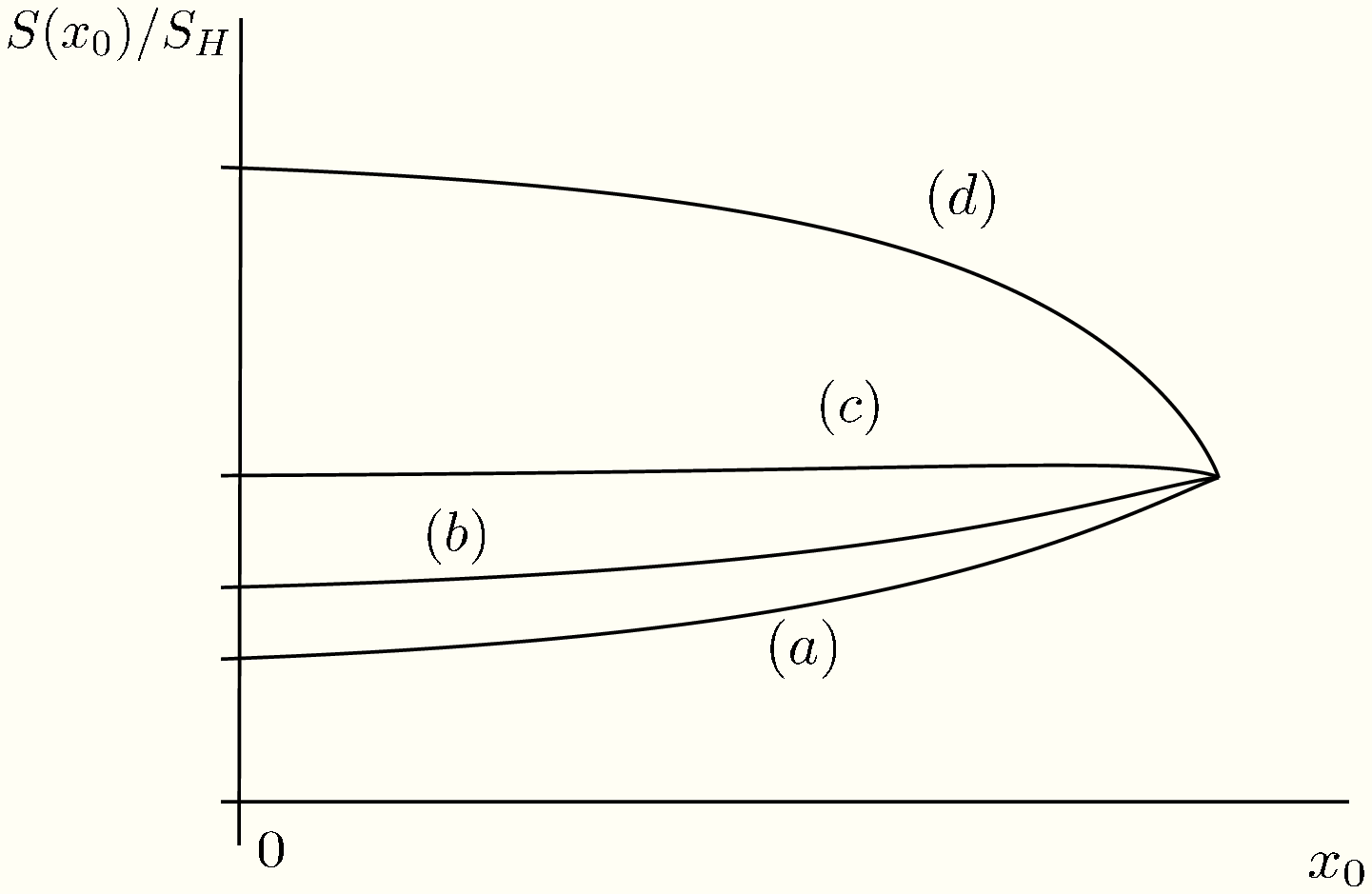}{4truein}

 The conclusion is that, asymptotically for $E_* \ll E < E_\Lambda$, the system is dominated by a large black brane
 with approximate Hagedorn spectrum. The upper limit marks the largest possible black brane that fits within the finite length of the tube. At the same time, any attempt at decoupling the tube cutoff, $x_\Lambda \rightarrow \infty$ at fixed $E$ and fixed $T_H$, results in $E_{\rm t} \rightarrow 0$ and $E_*, E_\Lambda \rightarrow \infty$, leaving behind {\it just} the phase of nonrelativistic radiation in the tube.

\subsec{Stability}

\noindent

It is interesting to study the stability properties of the combined system at the equilibrium point in the Hagedorn band.
A local entropy maximum occurs for high energies and moderate radial cutoffs, i.e.
$\shalf \log {\cal E} < x_\Lambda \sim N^\alpha {\cal E}$. In this case we have a local maximum for
$0<{\bar x} < \shalf \log {\cal E}$. The microcanonical inverse temperature at this maximum is
given by
$$
 \beta(E) = {dS \over  d E} = {\partial  S \over \partial x} {d{\bar x} \over dE }+
{\partial  S \over \partial  E} \;.
$$
At the local maximum $\partial_x  S =0$ and we are left with
$$
T_H \beta ( E) = {9\over 10} N^{-{\alpha\over 10}} \left({x_\Lambda - {\bar x} \over {\cal E} - e^{2{\bar x}}}\right)^{1\over 10}
$$
or, using the $\partial_x  S =0$ equation,
\eqn\mono{
T_H \beta ( E) = 1- {1\over 20} N^{-{\alpha\over 10}} \,e^{-2{\bar x}} \left(  {{\cal E} - e^{2{\bar x}} \over x_\Lambda -{\bar x}}\right)^{9\over 10}\;.
}
The correction term to the Hagedorn behaviour on the right hand side is negative
and decreases in size. Hence, we have a situation where $\beta( E)$ approaches unity from
below, resulting in a system with negative specific heat. For ${\cal E} > N^{\alpha} x_\Lambda$
and ${\bar x} \sim 0$, i.e. for the equilibrium at the beginning of the transient, we have corrections of
$O(1)$ to the Hagedorn temperature. For  larger ${\cal E}$, with ${\bar x} \approx \shalf \,\log \,{\cal E}$, we have smaller corrections.

In order to obtain a more accurate estimate of these small corrections we  must solve for ${\bar x}$ in the equilibrium equation
$\partial_x  S=0$, i.e.
$$
0= 2e^{2{\bar x}} - {1\over 10} N^{-{\alpha\over 10}} \,h^9 - {9\over 5} N^{-{\alpha\over 10}} \,e^{2{\bar x}} \,{1\over h}
\;,$$
where we denote
$$
h= \left({{\cal E} - e^{2{\bar x}} \over x_\Lambda -{\bar x}}\right)^{1\over10}
\;.$$
For ${\cal E} \gg 1$ and moderate values of $x_\Lambda$, we can approximate the equilibrium
equation by just the two terms proportional to $\exp(2{\bar x})$, so that
$$
h\approx  {9\over 10} N^{-{\alpha\over 10}}
\;.$$
Writing
$
{\bar x} = \shalf \log {\cal E} - \delta
$,
with $\delta \ll 1$ and substituting back into the definition of $h$ we obtain
$$
\delta \approx \half \left({9\over 10}\right)^{10} \, \left({x_\Lambda - \shalf \log {\cal E} \over N^\alpha {\cal E}}\right)
\;.$$
Going back to the general expression for the inverse temperature in units of $1/T_H$, we find
\eqn\stab{
T_H \beta ( E)  \approx 1-{1\over 20} \left({9\over 10}\right)^9 {1+2\delta  \over  N^\alpha {\cal E} }
\;,
}
with  the explicit dependence on the tube cutoff only showing up at second order in the large energy expansion.
We find at the end a result that resembles what was obtained in \refs\rkut. In particular, the ``critical exponent", $1+\gamma$, in front of the logarithmic
correction to the entropy is extensive in the world-volume of the fivebrane and scales like  it does in \entkut.
 Our derivation shows that the Hagedorn regime is indeed locally unstable, with negative specific heat, but the validity of the approximations involved requires that the tube cutoff is not too large compared with the energy, i.e. we have  $\shalf \log {\cal E} < x_\Lambda \ll
N^\alpha {\cal E}$.

 Therefore,
we find  a result consistent with the
analysis in the canonical ensemble: the Hagedorn behaviour is swamped by the one-loop corrections due to radiation in the tube, surviving only
in a finite band of energies associated to the high energy cutoff of the model. The monotonicity of \mono\ implies that the  specific heat
of this finite Hagedorn band is negative, with the possible exception of the immediate vicinity of
 the high energy cutoff, where the radiation effects are necessarily small. Indeed, the large-energy asymptotics    \stab\  is parametically dominated (by a factor of $N^4$) by the local loop corrections at the horizon, given by Eq. \col. 

\newsec{UV completion of  LST}

\noindent

Our main conclusion so far is that the classical Hagedorn phase of LST seems to be inextricably
associated to a high-energy cutoff that limits the extensivity of the `tube'.  On the other hand, both the microcanonical and the canonical analysis suggest that a first-order phase transition could take place, if a new set of {\it stable}  degrees of freedom  would take over at very high energies.  Thus, it is natural to try a physical realization of the cutoff that provides at the same time these stable degrees of freedom.
  The simplest possibility  would be
an embedding of the LST into an UV fixed point, i.e. some known CFT with an AdS description in the gravity regime.

If such an `ultraviolet completion' is feasible, we have a well-posed thermodynamical ensemble, with positive specific heat in the deep high energy regime, and stable under loop corrections, since radiation thermodynamics is perfectly compatible with holography in such spaces. At the same time,
one would be able to relate any possible instabilities of the intermediate `Hagedorn regime' to known
phenomena in the asymptotic CFT at high energies.

In this section we describe one such embedding, using the ideas of `deconstruction' \refs{\rdecons}, in which the world-volume of the NS5-branes is
partially compactified on an appropriate lattice. 
Various deconstructions of LST theories have been proposed \refs{\rmotl, \rdorey}. Here, we will follow the blueprint of \refs\rdorey\ for the type IIB LST (see also \refs{\radamsfab, \rdoreyp, \rjabbari}), and adapt it to the type IIA case. In type IIB theory, we consider an ${\cal N} =4$ super Yang--Mills model  with gauge group $U({\hat N})$ and a specific superpotential perturbation, called $\beta$-deformation:
$$
W = i\,\tr\left( e^{i{\bar \beta}/2} \Phi_1 \Phi_2 \Phi_3 - e^{-i{\bar \beta}/2} \Phi_1 \Phi_3 \Phi_2 \right)
\;,
$$
where $\Phi_i$ stand for the three chiral superfields of the ${\cal N} =4$ SYM theory. At the particular value ${\bar \beta} = 8\pi^2 i/g_{\rm YM}^2 n$, with $g_{\rm YM}$ the Yang--Mills coupling and $n$ a positive integer that divides ${\hat N}$, the theory has a moduli space of vacua with unbroken $U(N)$ gauge symmetry,
$N= {\hat N}/n$, and a spectrum  of massive modes that simulates a Kaluza--Klein reduction of a six-dimensional theory on an $n\times n$ lattice.

One way of visualizing this vacuum is in terms of the
S-dual configuration, in which the vacua in question furnish a standard Higgs branch for the $\beta$-deformed theory with ${\bar \beta}= 2\pi/n$. The particular configuration of the adjoint  fields at these vacua is non-commutative, with $\Phi_1$ and $\Phi_2$ proportional to the matrices ${\bf 1}_N \otimes \Gamma_1$ and ${\bf 1}_N \otimes \Gamma_2$ respectively,  and $\Gamma_{1,2}$ satisfying  $\Gamma_1 \Gamma_2 = \Gamma_2 \Gamma_1 \,\exp(-2\pi i/n)$,  the standard $n$-dimensional Weyl algebra of {\it clock} and {\it shift} matrices. Since the eigenvalues of $\Gamma_{1,2}$ are the $n$-th roots of unity, and the classical value of $\Phi_{1,2}$
give the D3-brane transverse positions in two complex planes,  the D3-branes can be seen as distributed in an $n\times n$ lattice torus. However, the given $\Phi_1$ and $\Phi_2$ do not commute, and the lattice is `fuzzy' or `noncommutative'.   For a dense set of D-branes, we may approximate this configuration using   the so-called Myers effect \refs\rmyers,  wrapping $N$ D5-branes on a torus that contains a uniform distribution of the ${\hat N}$ D3-branes. The D5-branes are marginally stabilized against shrinking
by the background flux turned on by the $\beta$-deformation, and the noncommutativity of the D5-branes is implemented by turning on an appropiate $B$-field flux. Expanding all fields around this vacuum configuration,  the standard Higgs mechanism yields a  massive spectrum that simulates a lattice compactification from a six-dimensional theory. 
 
 The D-brane construction has the advantage of determining the  dual geometry of this system. We have  an asymptotically ${\rm AdS}_5 \times {\bf S}^5$ throat,
 sourced by ${\hat N}$ D3-branes which are  smeared over a two-torus sitting at  radial coordinate $r=r_\varepsilon$. In the limit ${\hat N}\gg N\gg 1$ we can neglect the effect of the D5-branes until very near the smeared torus, where one enters a geometry of $N$ D5-branes. The  $B$-field profile is determined by its large-$r$ asymptotic value
$
\int_{{\bf T}^2} B_{\rm NS}  \rightarrow 2\pi \alpha' n$ (c.f. \refs{\rmaldarusso, \ritzhakihashimoto, \rozalisha}).

We are actually interested in the S-dual system, which does not admit a simple weak-coupling description in terms of a gauge-theory Higgs branch. However, the bulk geometrical description arising at large 't Hooft  coupling can be studied in either S-dual frames. 

\subsec{Gravitational dual of the deconstructed type IIB LST}

\noindent

By means of an S-duality of type IIB string theory, we can convert the background described in the previous subsection into
a deconstruction of the NS5-brane theory in type IIB. In this case, one has a system of ${\hat N}$ D3-branes, smeared over a torus sitting at $r=r_\varepsilon$, with $N$ wrapped NS5-branes. The $B$-field flux is now replaced by $n$ units of  Ramond--Ramond flux through the same torus.
This background is characterized by a standard `tube' regime at small radius, with intrinsic length scale $R$, and new features at   the radius where noncommutativity effects become important, $r_\theta$,  and the scale of the D3-brane smearing and NS5-brane wrapping, $r_\varepsilon$. Beyond this radius, we find an asymptotic AdS space with length scale $\CR_{\rm UV}$. We will see  that these parameters are constrained in the present set up, so that only a subset of them  remain independent.

The metric reads (c.f. \refs\rdorey)
\eqn\ncfb{
ds^2= \sqrt{1+r^2 /r_\theta^2 } \,(-dt^2 + dx_3^2) + {dv^2 + dw^2 \over \sqrt{1+r^2/ r_\theta^2}}
   + R^2 \sqrt{1+r^2/ r_\theta^2 } \, \left({dr^2 \over r^2} + d\Omega_3^2\right)\;,
}
and the dilaton profile
\eqn\dilp{
e^\phi = g_s \sqrt{1+ r_\theta^2 /r^2}\;.
}
This background  corresponds to $N$ noncommutative NS5-branes in the near-horizon limit with $R=\sqrt{N\alpha'}$. At $r\ll r_\theta$ we
recover \hololst\ with the $y$-coordinates split into $x_3 \in {\bf R}^3$ and $(v,w) \in {\bf T}_L^2$, a square
two-torus of size $L$. For $r \gg r_\theta$ we approach the metric of ${\hat N}=Nn$ D3-branes, smeared over the ${\bf T}^2_L$ at $r=r_\varepsilon \sim L$:
\eqn\smdt{
ds^2 = \left({\pi R_{{\rm D}3}^{\,4} \over L^2 r^2}\right)^{-1/2} \,(-dt^2 + dx_3^2 ) +
\left({\pi R_{{\rm D}3}^{\,4} \over L^2 r^2}\right)^{1/2} \left( dv^2 + dw^2 + dr^2 + r^2 \,d\Omega_3^2 \right)\;.
}
The matching with
\ncfb\ fixes $r_\theta = R$ and $R_{{\rm D}3}^2 = RL/ \sqrt{\pi}$. Defining the new radial coordinate
 $\rho = \sqrt{r^2 +v^2+ w^2}$ we find  the conformal ${\rm AdS}_5 \times {\bf S}^5$ metric with radius $\CR_{\rm UV} = R_{{\rm D}3}$,  for
 $\rho \gg r_\varepsilon \sim L$.

From the point of view of the CFT in the UV regime, the model is parametrized by two discrete parameters, ${\hat N}$ and $n$, and one continuous parameter, the 't Hooft coupling, ${\hat \lambda}$, in addition to the energy scale set by the particular vacuum chosen. We denote this energy scale as $\varepsilon^{-1}$ to signify its interpretation in the LST theory as a `lattice' scale. In the infrared regime,
the model flows to the rank-$N$ type IIB LST with lenght scale $R=2\pi T_H$. At intermediate scales,
the LST is compactified on a two-torus of size $L$ and has noncommutative features controled by the
$B$-field flux of $n$ units. The parametric mapping between the UV CFT and the noncommutative LST is obtained from the relation
\eqn\pmap{
{\hat \lambda} \sim {R_{{\rm D}3}^{\;4} \over \alpha'^{\,2}} \sim {R^2 L^2 \over R^4} N^2 \sim N^2 {L^2 \over R^2}\;.}
The main restriction of the present construction is the fact that the Hagedorn scale of the LST, $T_H$, is linked to the `lattice regularization' scale $\varepsilon^{-1}$. This follows from the UV/IR relation in the high-energy CFT, which gives
\eqn\uvirr{
\varepsilon^{-1} \sim {r_\varepsilon \over R_{{\rm D}3}^{\;2}} \sim {L \over RL} = {1\over R}\;.}
As explained in \refs\rdorey, strong-coupling effects in the gravity regime at ${\hat \lambda} \gg 1$ change the naive weak coupling interpretation of $n$ as the number of `lattice points' in a given direction. Using \pmap\ and \uvirr\ we can write for the `effective size' of the lattice,
\eqn\effla{
n_{\rm eff} \equiv {L \over \varepsilon} \sim {L \over R} \sim {\sqrt{{\hat \lambda}} \over N} = n\;{\sqrt{{\hat \lambda}} \over {\hat N}}\;.}
 The string coupling controls the length of the LST tube through the relation $x= \log (r/r_H)$, yielding
\eqn\gstring{
x_\theta = \log (N/g_s)\;.}
Hence, the LST with fixed $R$ and $N$ can be fitted with a long tube, $x_\theta \gg 1$, if the string coupling $g_s$ is sufficiently small. The requirement of a `large lattice', $n_{\rm eff} \gg 1$ is met by
choosing $n\gg N$, since $n_{\rm eff} \sim \sqrt{\hat \lambda} /N \sim \sqrt{g_s n/N}$.

Finally, it is useful to express the important thresholds of the system in terms of the energy attained by a black brane with horizon reaching out to $r_H, r_\theta$ and $r_\varepsilon$. The first is the Hagedorn energy, i.e. the minimum energy for which the Bekenstein--Hawking entropy is Hagedorn-like,
\eqn\hagen{
E_H \sim N^2 V_3 \, L^2 \,T_H^{\;6}\;.}
Next we have the the energy threshold at the end of the (commutative) tube,
\eqn\endt{
E_\theta = E_H \,e^{2x_\theta}\;,}
and finally the energy of the black brane at the `deconstruction scale'. We can define it in terms of the UV parameters as the energy of a black D3-brane with horizon at $r_0 \sim r_\varepsilon \sim L$.
\eqn\decen{
E_\varepsilon \sim {\hat N}^2 \, V_3 \,T_H^{\,4}\;.}
Using the previous expressions, we can obtain the following hierarchies:
\eqn\hierar{
{E_\theta \over E_H} \sim {n^2 \over n_{\rm eff}^{\,4}}\;, \qquad {E_{\varepsilon} \over E_H} \sim {n^2 \over n_{\rm eff}^{\,2}}\;,}
and, in particular $E_\varepsilon / E_\theta \sim n_{\rm eff}^{\,2}$. Hence, a long LST tube, with $x_\Lambda \gg 1$ at fixed $N$, requires $n \gg n_{\rm eff}^{\,2}$, whereas a large deconstruction lattice requires $n_{\rm eff} \gg 1$, resulting in a large band of energies with noncommutative LST behaviour. We can supress the noncommutative band by using a small deconstruction lattice, $L\sim \varepsilon \sim R$, while still having a weakly curved and weakly coupled background provided $n\gg N \gg 1$. The price we pay for removing the noncommutative band is that there is no regime in which the  LST is effectively six-dimensional, as $LT_H \sim 1$.

\subsec{Gravitational dual of the deconstructed type IIA LST}

\noindent

The type IIA case follows from the type IIB one by a T-duality along one of the compact circles, say the
$v$ coordinate. In the deep infrared, $r\ll r_\theta$, the NS5-brane metric is self-dual, whereas we recover at $r\gg r_\theta$ the metric of ${\hat N}$ D4-branes smeared over
 the $w$-circle and wrapped on the $v$ circle. In fact, we may as well decompactify the $v$ circle and let the $x_4 = (x_3, v)$ coordinates parametrize
the spatial ${\bf R}^4$ factor of the D4-branes worldvolume. We obtain:
\eqn\ncfba{
ds^2 =\sqrt{1+r^2 /r_\theta^2} \,(-dt^2 + dx_4^2) + {dw^2 \over \sqrt{1+r^2 /r_\theta^2}} +
R^2 \sqrt{1+r^2 /r_\theta^2} \;\left({dr^2 \over r^2} + d\Omega_3^2 \right)\;.
}
The dilaton is given by the same expression as in the type IIB case, eq. \dilp, with $g_s$ denoting now the type IIA string coupling, whereas the $n$ units of RR flux through the $(v,w)$ torus is T-dualized into
a Wilson line of RR one-form through the $w$-circle. Matching the metric \ncfba\  to that of ${\hat N}$ D4-branes smeared over a circle
of size $L$,
\eqn\smdf{
ds^2 = \left({Lr^2 \over 2R_{{\rm D}4}^{\,3}}\right)^{1/2} \,\left(-dt^2 + dx_4^2 \right) + \left({2R_{{\rm D}4}^{\,3} \over L r^2}\right)^{1/2} \,\left(dw^2 + dr^2 + r^2 \,d\Omega_3^{\,2} \right)\;,
}
we obtain $2 R_{{\rm D}4}^{\,3} = R^2 L$ and $r_\theta = R $. Defining again an asymptotic radial coordinate
 $\rho = \sqrt{r^2 + w^2}$ the effect of the smearing disappears for $\rho \gg r_\varepsilon \sim L$, and
we converge to the metric of ${\hat N}$ localized D4-branes,
\eqn\dcuatro{
ds^2 = \left({\rho \over R_{{\rm D}4}}\right)^{3/2} \left(-dt^2 + dx_4^2 \right) + \left({R_{{\rm D}4} \over \rho}\right)^{3/2} \,\left(d\rho^2 + \rho^2 \,d\Omega_4^2\right)\;.
}
Unlike the type IIB case, this metric does not provide a full UV description by itself. At a radius of order
$\rho \sim R_{{\rm D}4} g_s^{-1/4}$, the dilaton is of $O(1)$ and we must uplift the metric to the near-horizon regime of
a compactified stack of ${\hat N}$ M5-branes in eleven-dimensional supergravity. The resulting asymptotic geometry
is ${\rm AdS}_7 \times {\bf S}^4$ with radius $\CR_{\rm UV} =  2\ell_p \,(\pi {\hat N})^{1/3}$, with $\ell_p$ the eleven-dimensional Planck length $\ell_p = g_s^{1/3} \ell_s$.

As in the IIB case, the deconstruction scale, $\varepsilon^{-1}$, is tied to the Hagedorn temperature of the LST.
The UV/IR relation of the five-dimensional $U({\hat N})$ Yang--Mills theory on the D4-branes yields $\varepsilon \sim \sqrt{R_{{\rm D}4}^{\,3} /r_\varepsilon} \sim R \sim \beta_H$. Using the general relation $R_{{\rm D}4}^{\,3} \sim \ell_s^3 \,g_s {\hat N}$ we derive for the effective lattice spacing
\eqn\latef{
n_{\rm eff} = {L\over \varepsilon} \sim {L\over R} \sim {g_s n \over \sqrt{N}}
\;.}
On the other hand, the relation between the commutative tube length and the string coupling becomes $x_\theta \sim \log(\sqrt{N}/g_s)$.
Finally, the LST threshold energies are given by
\eqn\hagdosa{
E_H \sim N^3 V_4 \,L\,T_H^{\,6}\;, \qquad E_\theta = E_H \,e^{2x_\theta}\;,
}
whereas
\eqn\deconenn{
 E_\varepsilon \sim {\hat N}^2 \,(g_s {\hat N} \ell_s) \,V_4 \,T_H^{\;6}
 }
 stands for the energy of black D4-branes with horizon at the deconstruction radius $r_0 \sim r_\varepsilon$.
 Putting all together, we find the following relations for the energy hierarchies:
\eqn\morehi{
{E_\theta \over E_H} \sim {n^2 \over n_{\rm eff}^{\,2}} \;,\qquad {E_\varepsilon \over E_H} \sim n^2\;.}
The size of the noncommtative band is controlled by $n_{\rm eff}$, just as in the type IIB case, i.e.  $E_\varepsilon \sim n_{\rm eff}^{\,2} \, E_\theta$.

\fig{\sl Schematic picture  of the background profile in IIA deconstruction, showing the different regions of interest in the vicinity of the LST regime. For the type IIB case, we replace the low-radius  ${\rm AdS}$ by a weak-coupling wall and the ${\rm D4}$-branes by ${\rm D3}$-branes.
}{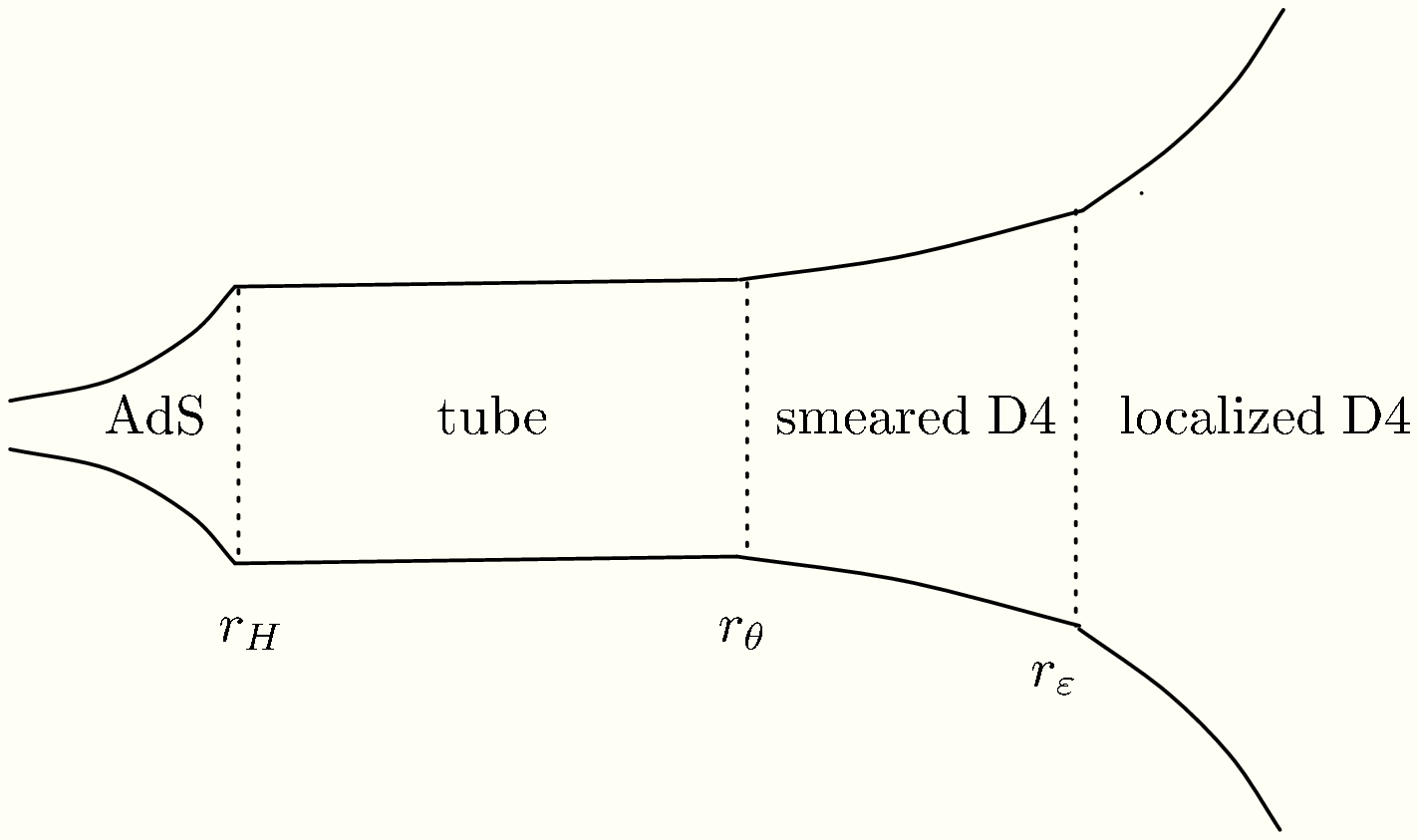}{3.5truein}

We may summarize the energy hierarchies for both IIA ($\alpha =3$) and IIB ($\alpha=2$)  cases by the relations
\eqn\hiero{
{E_\theta \over E_H} \sim \left({n \over n_{\rm eff}^{4-\alpha} }\right)^2 \;, \qquad {E_\varepsilon \over E_H} \sim \left({n \over n_{\rm eff}^{3-\alpha}}\right)^2\;.}

\newsec{Thermodynamics of the deconstructed LST theories}

\noindent

The Hagedorn phase of the LST theories is recovered in the gravitational dual by considering the
corresponding near-extremal black metrics, obtained by the substitution  $dt^2 \rightarrow dt^2 h(r)$ and $dr^2 \rightarrow dr^2 /h(r)$, with $h(r) = 1-r_0^2 /r^2$ for all the noncommutative NS5-brane
metrics or the smeared metrics at large radius \refs\rozalisha. 

It is important to notice, however, that these metrics are only approximate, as the full background should not admit strictly static solutions at finite temperature. Notice that the  mass scale in the deconstructed LST theories comes from a choice of a particular vacuum in the Higgs branch of the $\beta$-deformed CFT. At finite temperture, one expects these flat directions to be lifted by thermal effects. The `lifetime' of the LST vacuum
with respect to this process can be made arbitrarily slow by increasing the `rest mass' of the branes, i.e. by
the $n\rightarrow \infty$ limit. In the gravitational dual, this limit is precisely corresponding to making the `tube' arbitrarily long. Since the time dependence is related to the instability of the wrapping torus at radius $r_\varepsilon$, we expect that near-extremal solutions with $r_0 \ll r_\varepsilon$ will be approximately static. Indeed, exact solutions for a static non-extremal black brane exist for an infinite tube, both in the commutative and non-commutative cases.  
As $r_0$ approaches the deconstruction scale $r_\varepsilon$, the dynamical instability becomes more important, and our approximate picture based on static solutions is more qualitative. In effect, when the horizon rises past $r_\varepsilon$, the smeared black-brane metrics characteristic of the Higgs branch become 
locally unstable towards localization on the two-torus (via the Gregory--Laflamme tachyonic mode).

Thus, in the deep ultraviolet regime we describe the system in terms of the normal vacuum of the high energy theory, with  $h(\rho)= 1-\rho_0^4 /\rho^4$ in the type IIB case,  corresponding to a fully localized D3-brane stack, and $h(\rho) = 1-\rho_0^3 /\rho^3$ in the type IIA case, as appropriate for the localized D4-brane stack. 
We effectively  deal with the global thermal instability of the LST construction by taking the static aproximation for $r_0 \ll r_\varepsilon$ and switching to the localized solution abruptly at $r_0 \sim r_\varepsilon$, a procedure that we expect to be qualitatively sensible for $n\gg 1$. 

Under these assumptions,
we end up with metrics of the form
$$
ds^2 = -F(r)dt^2 + {dr^2 \over F(r)} + \dots,
$$
 for which the Hawking temperature is given by $T(r_0) = F'(r_0)/4\pi$. Such a temperature is invariant under a conformal rescaling of $F(r)$, provided it is non-singular at the horizon $r=r_0$. As a result, the effect of noncommutativity, given by the factors $\sqrt{1+r^2/r_\theta^2}$ in the five-brane metrics, is inocuous for the calculation of the  Hawking temperature and the Hagedorn plateau $T(r_0) = T_H =
(2\pi R)^{-1}$ persists until we reach $r\sim r_\varepsilon \sim L$. Beyond this point, we enter the standard behaviour of D3-branes, with $T(\rho_0) R \sim \rho_0 / L$,  in the type IIB case. Conversely, in the type IIA case one finds the  D4-brane temperature curve, $T(\rho_0) R\sim\sqrt{\rho_0 /L}$,  followed by that of M5-branes, $T(\rho_0) R \sim \rho_0 / RL$, in the asymptotic eleven-dimensional regime. In all these cases, we have
a normal finite-temperature ensemble over a particular vacuum of a dual CFT in $d$ spacetime dimensions. We expect then that
the transition to the standard positive specific heat regime $T\sim E^{1/d}$ will take place smoothly. Hence, the classical approximation to $T(E)$ yields a Hagedorn plateau which tilts upwards for
$r_0 > r_\varepsilon \sim L$, with positive specific heat (c.f. Fig 10).

\subsec{Radiation corrections}

\noindent

We are now ready to estimate the low-energy  radiation contribution to the one-loop thermodynamics. As emphasized above, this is the possible source of extensive  behaviour in the holographic coordinate.
The deconstructed models feature an LST tube extending over the region $z_H \ll z \ll z_\theta$, with
$z_H = R \log(N^{-2 + \alpha/2})$ and $z_\theta = R\log(1/g_s)$. The radiation thermodynamics of the tube is as explained in Section 5, i.e. we have a nonrelativistic seven-dimensional gas for $T\ll T_H$ and a relativistic ten-dimensional gas for $T\gg T_H$.

Further up the throat, in the region $z_\theta \ll z\ll z_\varepsilon$, with $z_\varepsilon =R \log (L/Rg_s)$,
we find the noncommutative NS5 regime, with an associated effective potential
\eqn\effnc{
V_{\rm eff} (z) \approx {\vec p}_c^{\,2} + f(z)^2 \,{\vec q}^{\;2} + {\Delta_{{\bf S}^3} \over R^2} +\left({\alpha\over 2R}\right)^2
\;,}
where ${\vec p}_c$ stands for the momentum in the $\alpha +1$ commutative directions and ${\vec q}$
is the momentum in the noncommutative torus of size $L$ and dimension $4-\alpha$. The potential is qualitatively very different depending on whether we excite the momentum modes, ${\vec q}$,  in the noncommutative directions. For $z\gg z_\theta$, we have $f(z) \approx \exp(z/R)$, inducing a steep wall in the $z$ direction, except for ${\vec q}=0$, for which we have a mass gap $\alpha^2 /4R^2$.  We can then evaluate \resl\ for this zero-mode sector, by integrating momenta {\it only} along the commutative directions:
\eqn\vzmm{
\Lambda(\tau_2)_{{\vec q}=0} =\sum_{\Delta_{{\bf S}^3}} {(-1)^F  \over (4\pi^2 \alpha' \tau_2)^{\alpha +3 \over 2}} \; V_5 \,L^{\alpha-4} \, \Delta z\;\exp\left(-\pi\alpha' \tau_2 \left({\Delta_{{\bf S}^3} \over R^2} + {\alpha^2 \over 4 R^2}\right)\right)\;,}
where we have assumed $z\gg z_\theta$. We see that the radiation is extensive in the volume of the commutative directions, $V_5 /L^{4-\alpha}$, as well as the $z$ direction, i.e. it behaves as a {\it nonrelativistic}  $(\alpha+3)$-dimensional gas for $T\ll T_H$. For $T\gg T_H$ we activate the extensivity in ${\bf S}^3$, leading to an $(\alpha+6)$-dimensional {\it relativistic} gas.

On the other hand, the contribution of the higher momentum modes in the noncommutative directions scales with the optical volume of the spatial metric,
\eqn\vnzm{
\Lambda(\tau_2)_{{\vec q}\neq 0} \approx \sum_{\Delta_{{\bf S}^3}} {{(-1)^F} \over (4\pi^2 \alpha' \tau_2)^{7/2}} \;V_5\; \int_{z_\theta}^{z_\varepsilon} {dz \over f(z)^{4-\alpha}} \;\exp\left(-\pi\alpha' \tau_2 \left({\Delta_{{\bf S}^3} \over R^2} + {\alpha^2 \over 4 R^2}\right)\right)\;.}
In this case, since the integral over $z$ is of $O(R)$, we have six-dimensional  nonrelativistic extensivity at $T\ll T_H$ and nine-dimensional relativistic extensivity for $T\gg T_H$.\foot{This contribution was estimated in \refs\ncen, where the lower-dimensional zero-mode term \vzmm\ was neglected, as the authors were focusing  on  entropy densities in the thermodynamic limit.}

\fig{\sl Effective radiation potential in the deconstructed type IIA LST background. For $z_\theta \ll z\ll z_\varepsilon$ the tube plateau steps up to $(3/2R)^2$ for modes with zero momentum in the noncommutative direction, whereas we encounter an exponential wall for non-vanishing momentum modes.
}{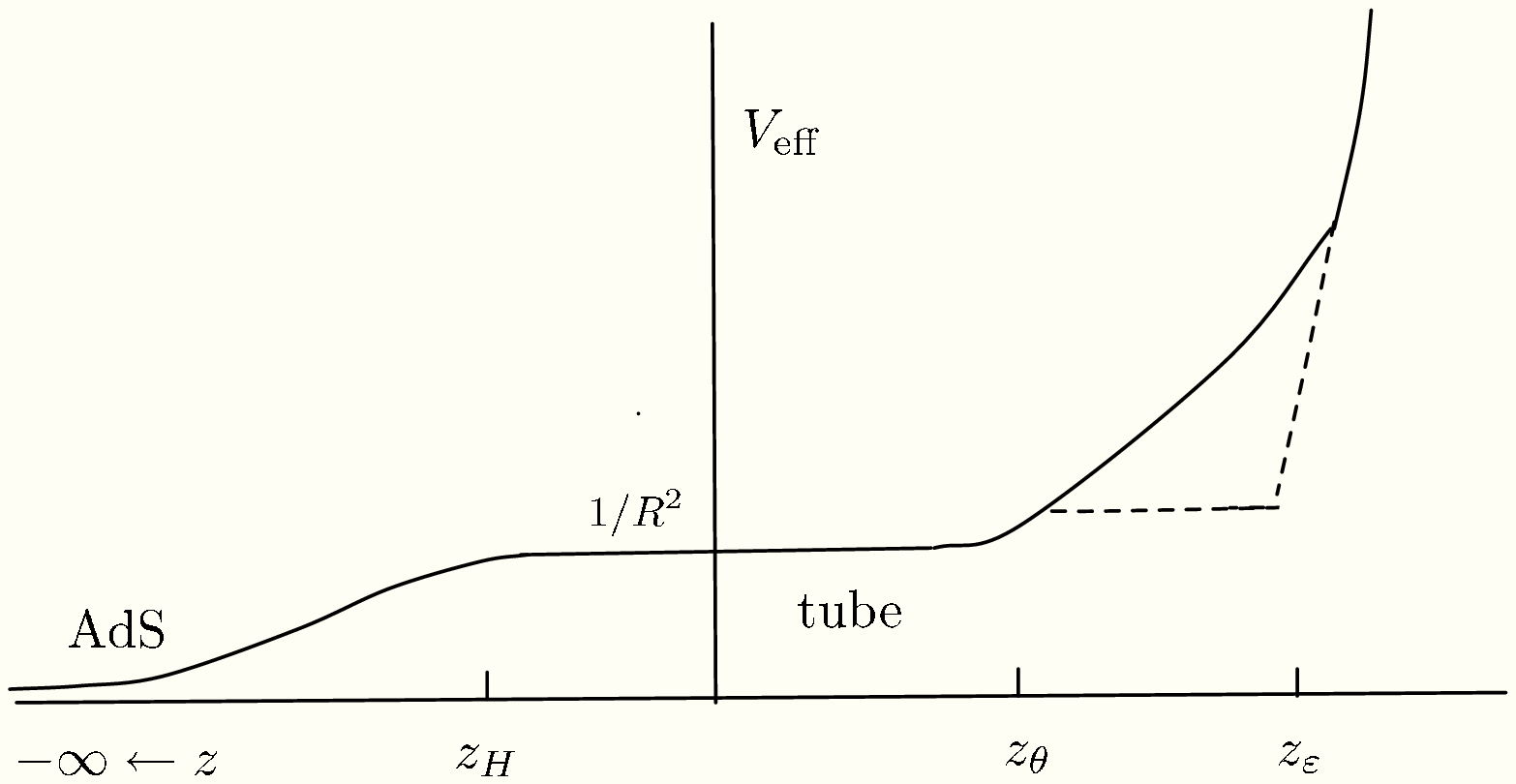}{4truein}

Hence, for  $z_\varepsilon \gg z_\theta$, the dominant contribution in the noncommutative region  comes from the ${\vec q}=0$ sector \vzmm,   a radiation gas with similar properties to that in the tube plateau, except for its lower  effective dimensionality. In this case, the effective radiation cutoff is $x_\Lambda \sim x_\varepsilon$ and $E_\Lambda \sim E_\varepsilon$, although the details of the analysis of Section 5 must be slightly corrected for black energies in excess of $E_\theta$, because of this lower effective dimension of the thermal gas.
If $n_{\rm eff} \sim 1$, so that the LST tube merges directly with the field-theoretical UV completion, we can set $E_\Lambda \sim E_\theta \sim E_\varepsilon$. In this case, we also have $x_\theta \sim x_\Lambda$. 

\fig{\sl Phase diagram of the deconstructed LST theory, as a function of the energy and the effective length of radiation tube, $x_\varepsilon$, in the classical approximation. The noncommutative LST transient is supported by the ratio $n_{\rm eff} = L/\varepsilon \gg 1$, disappearing when it approaches unity.}{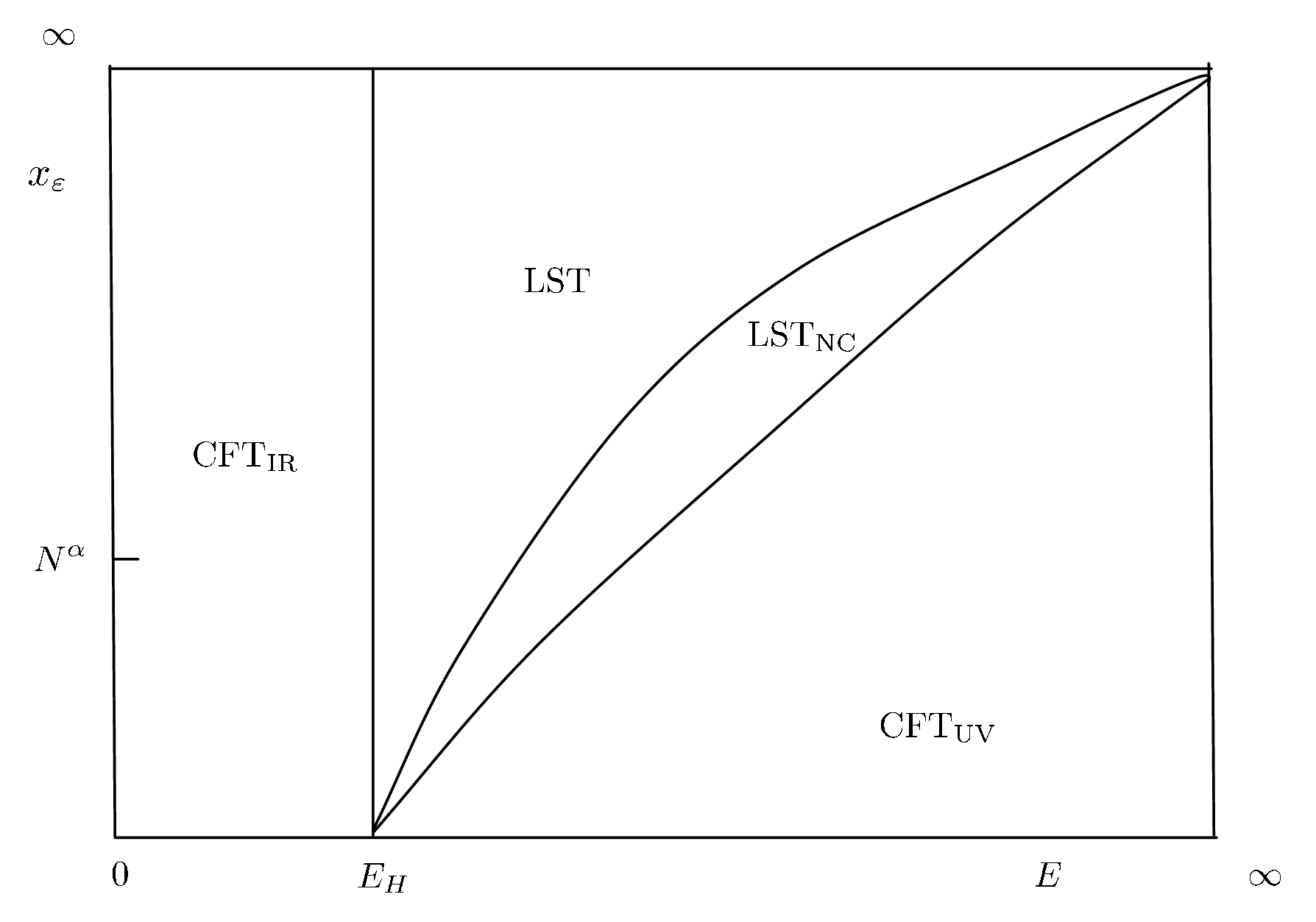}{4truein}

\subsec{Phase structure}

\noindent

We can now put all the previous results together and  plot the density of states or, equivalently, the microcanonical temperature as a function of energy
$T(E) = (\pt S /\pt E)^{-1}$, (c.f. Figs 8, 9, 10). 
 The deconstructed LSTs can be  parametrized by $N$, $T_H$, the total tube length $x_\varepsilon$, and the partial compactification length $L$. At the level of classical supergravity, the
thermodynamics only depends on two thresholds: $E_H$, determining the access to the Hagedorn plateau, and $E_\varepsilon$, determining the end of the plateau and the onset of the field-theoretical density of states of the UV completion.

The picture changes  when we add the one-loop corrections, in the form of radiation effects. Three new thresholds appear: $E_{\rm t}$ \etbe, determining the beginning of radiation domination in the tube, $E_*$, \bthr\ the postponed new door to the Hagedorn regime, and $E_\theta$,  \endt\ the threshold of noncommutative effects.

\fig{\sl The complete phase diagram of the deconstructed LST theory, as a function of the energy and the tube cutoff, $x_\varepsilon$. Comparing with the phase diagram in the classical approximation, we see that the main effect of the radiation corrections is to open a wedge for large tubes, $x_\varepsilon \gg N^\alpha$, dominated by the  seven-dimensional nonrelativistic gas.}{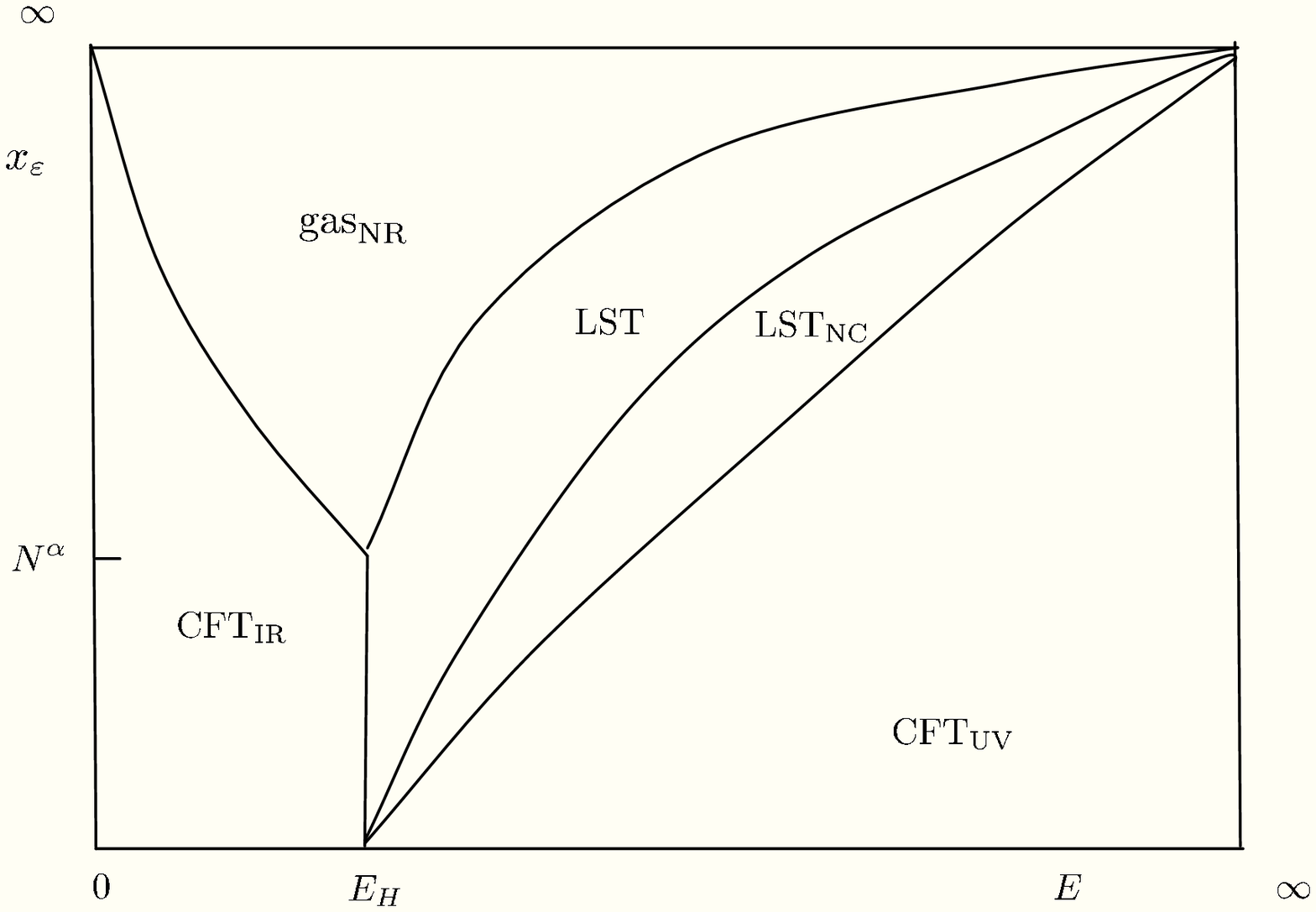}{4truein}

The function $T(E)$ shows six-dimensional CFT scaling  for $E<E_{\rm t}$. In the transient $E_{\rm t} < E< E_*$, the behaviour is that of nonrelativistic seven-dimensional gas. For $E_* < E<E_\theta$ we have
an approximate Hagedorn plateau with negative specific heat, which decays slightly faster  for $E_\theta < E < E_\varepsilon$, due to the lower dimensionality of the radiation in this region. Finally, the function $T(E)$ rises  for  $E>E_\varepsilon$ towards the UV fixed points, dominated by the classical contribution.

\newsec{Concluding remarks}

\noindent 

The curve $T(E)$ depicted  in Fig. 10 shows that the system undergoes a first-order phase transition when temperatures are rised above   $T_*$ \limt,  jumping directly from the phase of a nonrelativistic gas in the tube to the plasma phase
of the CFT in the UV. This means that the Hagedorn phase is an unstable  superheated transient behind the plasma phase of the asymptotic theory.   Hence, the situation is very similar to that of critical ten-dimensional Hagedorn ensembles, with the first-order phase transition at $T\sim T_*$ being similar to the Hawking--Page transition in AdS space (c.f. \refs\radshag), except that in this case the transition is triggered quantum mechanically in the bulk dynamics, and is not detected classically.

\fig{\sl The microcanonical temperature function, showing stable, unstable and metastable  phases, as a function of the important thresholds in the deconstructed LST systems. The dashed line represents the classical approximation, with a Hagedorn plateau extending from $E_H$ out to $E_\varepsilon$.  The dotted line gives the Maxwell construction of the first-order phase transition at the critical temperature $T_*$.  In the limit when we decouple the UV fixed point, $E_\varepsilon \rightarrow \infty$,
the high-energy thresholds of the tube $E_*$ and $E_\theta$  diverge, whereas the low-energy threshold $E_{\rm t}$ vanishes. In this limit, only the nonrelativistic gas remains.}{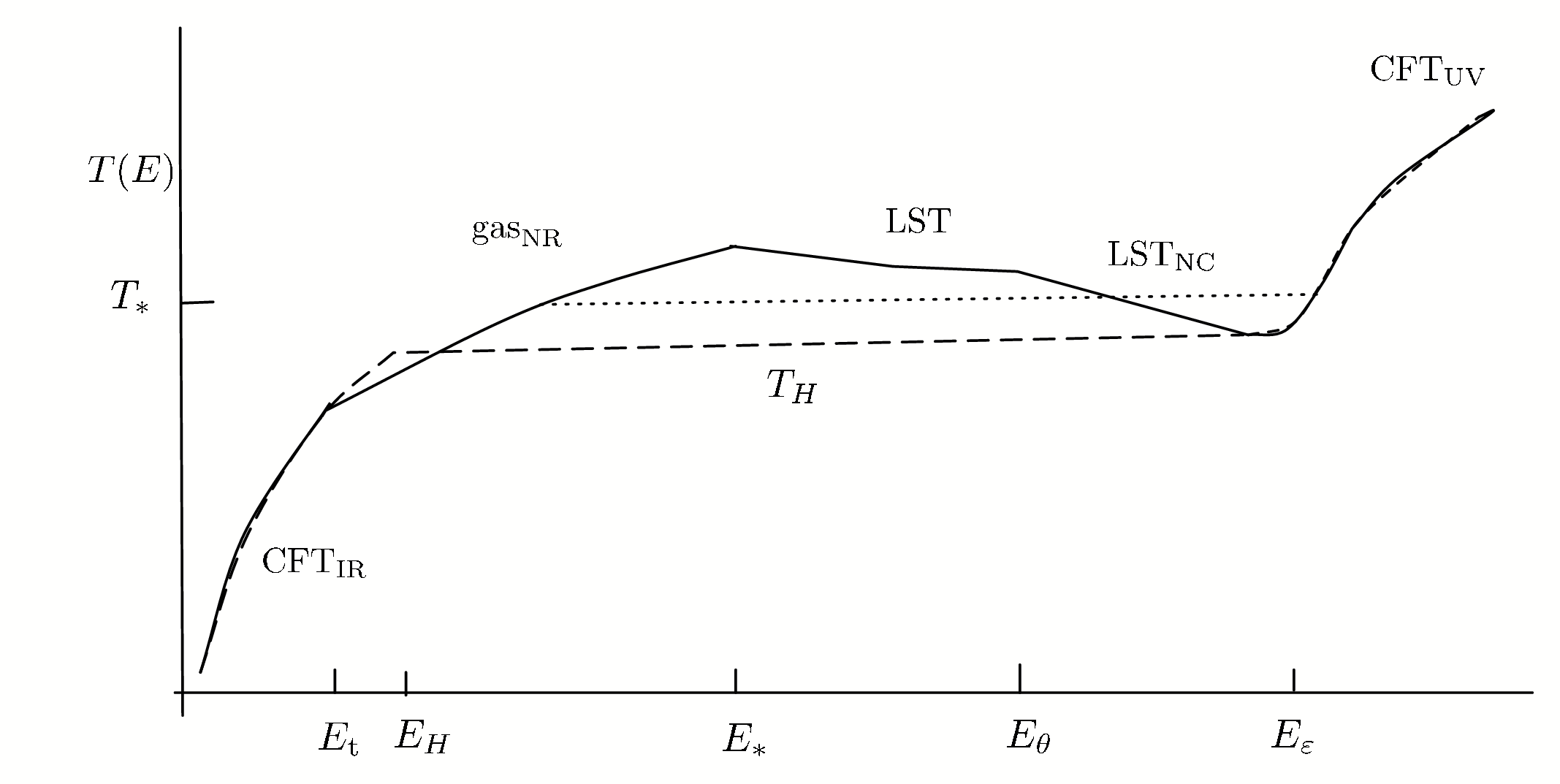}{4.5truein}

The dynamical relevance of the continuum of states above the tube gap was already emphasized in \refs\rlszlst, where it was found that these states contribute with zero-momentum singularities (hence infrared effects) in correlation functions of LST. Here we have seen that they even dominate the high-energy density of states,  potentially screening all characteristic `Hagedorn' effects of the LST. 

From the point of view of defining the LST's density of states, we confirm the results of the previous section, namely the lattice regularization cannot be removed without taking away the Hagedorn plateau with it, in effect  `throwing the baby out with the bath water'. In fact, not even the six-dimensional CFT behaviour at low energies remains, since $E_{\rm t} \rightarrow 0$ as $x_\varepsilon \rightarrow \infty$. In the decoupling limit we are left with the seven-dimensional nonrelativistic gas. It is important to emphasize, however, that the temperature of this nonrelativistic gas on the semi-infinite tube can never rise above $T_H$. If one tries to do so, the system is no longer in equilibrium and the  black brane proceeds to  engulf the whole tube, a statement that we made precise in the models with a tube cutoff, with $T_H$ replaced by $T_*$.  

If the tube is cutoff to a finite length, a Hagedorn band of (black-brane)  states can be identified at energies close to the cutoff, with a marginally negative specific heat. It is thus natural to embed the model into a larger system with stable high-energy thermodynamics, such as an asymptotic CFT in the UV. The tube cutoff appears then as a threshold separating the LST dynamics from the standard CFT dynamics at very high energy. 
 In this situation, one faces the following dicotomy: the model can be defined at sub-critical temperatures,  $T\leq T_H$, in such a way that  the dynamics decouples from the threshold, but showing no signs of a Hagedorn spectrum. Alternatively, one can give up decoupling,  having a Hagedorn spectrum that survives as a `threshold effect' at energies
close to the LST effective cutoff scale.   

The fate of the   LST Hagedorn transition, as discussed in this  paper, resembles that of the standard ten-dimensional critical strings  \refs\radshag\ in one important aspect: when the system is forced at temperatures above $T_H$, it reacts by tearing appart the spacetime structure (metrical and topological)  that supports the Hagedorn degrees of freedom (c.f. \refs\rtouring), i.e.  long strings in the critical case, or a horizon embedded in a tube in the LST case.

\vskip 1cm

{\centerline {\bf Acknowledgements}}

\noindent

We thank  M. Berkooz, S. Elitzur, D. Kutasov, N. Seiberg and especially O. Aharony and N. Dorey for a very useful correspondence. 
 The work of J.L.F.B. was partially supported by MEC
 and FEDER under grant
FPA2006-05485, CAM under grant HEPHACOS P-ESP-00346 and
 the European Union Marie Curie RTN network under contract 
 MRTN-CT-2004-005104. The work of C.A.F. was partially supported by MEC under FPU grant AP2005-0134.
The work of E.R. was partially supported by European Union Marie Curie 
 RTN network under contract  MRTN-CT-2004-512194, the American-Israel Bi-National Science Foundation, the Israel Science Foundation, The Einstein
 Center in the Hebrew University and by a grant of DIP (H.52).

\appendix{A}{WKB estimates of field-theoretical traces in curved spacetime}

\noindent

In this appendix we show how the principle of local extensivity in terms of red-shifted temperatures follows from a WKB approximation to the basic free energy trace
\eqn\freeet{
I_r = \Tr (-1)^F \,\log\left(1-(-1)^F e^{-\beta \omega}\right)\;.}
Let us consider, in general, traces of the form
\eqn\trf{
\Tr F(\beta \,\omega) = \sum_{n,j} F(\beta\,\omega_{n,j})\;,}
where $\omega^2_{n,j}$ is the eigenvalue spectrum of a family of positive-definite Schr\"odinger operators,
$ -\pt_z^2 + V_j ( z)$, indexed by $j$. Let us approximate each trace at fixed $j$
by  an integral
\eqn\asi{
\sum_n F(\beta\,\omega_n) \approx \int dn F(\beta\,\omega_n)
 = -\int n(\omega) \,dF(\beta\,\omega)\,,}
 where we have integrated by parts in the last step and neglected endpoint contributions. This expression  can be further approximated using the WKB relation
 \eqn\wkb{
 n (\omega)  \approx {1\over \pi} \int_{z_-}^{z_+} dz \,\sqrt{\omega^2 - V (z)}\;,}
where the integral over $z$ runs between the turning points of the given eigenstate and we assume $n\gg 1$.  Hence, we have the WKB evaluation
\eqn\wkbe{
\Tr \,F(\beta\omega) \approx -{1\over \beta \pi} \sum_j \int dz\,dy \,F'(y) \,\sqrt{y^2 - \beta^2 V_j (z)}\;,}
where the limits on the integrals and the $j$ sums are determined by the positivity of the square root in the integrand.

We may now apply \wkbe\ to the evaluation of \freeet, with $V_j (z)$ given by the effective potential \effpot. We obtain
\eqn\ffu{
I_r \approx -{1\over \beta\pi} \sum_{\Delta_i} \int dz dy  {\sqrt{y^2 - \beta^2 V_{\Delta_i} (z)} \over e^{y} - (-1)^F}\;.}
In some cases, it is useful to further approximate the sum over Laplace spectra $\Delta_i$ by integrals over the warped manifolds with metrics $ds_i^2$ in \genmrw. The condition for this to be a good approximation is that the local red-shifted temperature $\beta(z)^{-1} \equiv (\beta \sqrt{f(z)})^{-1}$be larger than the inverse typical `size' of the manifold, i.e.
\eqn\condc{
\beta(z) \,\rho_i (z) L_i \ll 1\;.}
Under these circumstances, we can approximate
\eqn\sumdel{\eqalign{
\sum_{\Delta_i} G\left(\beta(z)^2 \sum_i \Delta_i / \rho_i^2\right)
\approx & \prod_i {V_i \over (2\pi)^{d_i}} \int d^{d_i} {\vec k}_i \;G\left( \sum_i \beta(z)^2 {\vec k}_i^{\,2} /\rho_i^2\right) \cr= &\shalf |{\bf S}^{D-3}| \,\prod_i \left({V_i \rho_i^{d_i} \over (2\pi \beta(z))^{d_i}} \right) \;\int d\kappa \,\kappa^{{D\over 2}-2} \;G(\kappa) \;,}}
for a general function $G$. In the particular case of \ffu\ we finaly obtain
\eqn\finrs{
I_r \approx \int d{\vec V} \;\,\CI\left[\,\beta(z), M_{\rm eff} (z)\,\right]\;,}
where $d{\vec V} = dz \sqrt{f(z)} \prod_i \rho_i^{d_i} dV_i$ and  $\CI (\beta, M)$ is the flat-space free energy density for a single degree of freedom of mass $M$ at temperature $\beta^{-1}$, i.e.
\eqn\flats{
\CI(\beta, M) = -{1\over (4\pi)^{D-1 \over 2}} {\beta^{1-D} \over \Gamma\left({D+1 \over 2}\right)} \int_{\beta M}^\infty  dy\;{\left(y^2 - \beta^2 M^2\right)^{D-1 \over 2} \over e^y - (-1)^F}\;.}
The position-dependent mass and temperature are  defined by
\eqn\effm{
M_{\rm eff} (z)^2 = {V_0 (z) \over f(z)}\;, \qquad {1\over \beta(z)} = {1\over \beta \sqrt{f(z)}}\;,}
where $V_0 (z)$ is the  `s-wave' effective potential, i.e.  \effpot\ evaluated at $\Delta_i =0$. Notice that \effm\ equals ${\widetilde m}$ in \effpot, except for the back-scattering terms
that depend on $\rho_\Pi$. Hence, we have obtained the principle of local extensivity from the WKB approximation to the partition sum, as well as the more important corrections to it. 

Alternatively, we can start from the proper-time representation of \freeet:
\eqn\ptime{
I_r = -{\beta \over 4\sqrt{\pi}} \int_0^\infty {ds \over s^{3/2}} \sum_{\ell \in {\bf Z}} (-1)^{(\ell+1)F} \;e^{-{\ell^2 \beta^2 \over 4s}} \;\Tr \;e^{-s\,\omega^2}\;,}
and perform the WKB approximation to the trace, with $F(\beta\omega) =\exp(-s\omega)$,  to obtain
\eqn\wkbpt{
\Lambda (s) = {(-1)^F \over \sqrt{4\pi s} }\, \Tr \;e^{-s\,\omega^2} = {(-1)^F \over 4\pi s} \sum_j \int dz \;e^{-s\;V_j (z)}\;,}
an expresion that leads directly to \resl\ under $s=\pi\alpha' \tau_2$. If we may further approximate
the sum over $\Delta_i$ by  an integral, using \sumdel:
\eqn\finalpt{
\Lambda(\tau_2) \approx {(-1)^F \over (4\pi^2 \alpha' \tau_2)^{D/2}} \;\int d{\vec V}_{\rm op} \;e^{-\pi\alpha' \tau_2 \,V_0 (z)}\;,}
where $d{\vec V}_{\rm op}$ is the spatial volume element in the optical metric:
\eqn\opmet{
ds^2_{\rm op} = -dt^2 + dz^2 + f(z)^{-1} \sum_i \rho_i (z)^2 \,ds_i^2\;.}
Hence, we have also shown that radiation thermodynamics is linearly extensive in the volume of the optical metric.

\listrefs

\bye